\documentclass{aa}  

\usepackage{graphicx}
\usepackage{txfonts}
\usepackage{physics}
\usepackage[none]{hyphenat}
\begin{document} 

   \newcommand{\unit}[1]{\ensuremath{\, \mathrm{#1}}}
   \newcommand{\ramses}{{\sc ramses}}          % code names
   \newcommand{\krome}{{\sc krome}}          % code names
   \newcommand{\vapor}{{\sc vapor}}          % code names
   \newcommand{\dispatch}{{\sc dispatch}}          % code names
   \newcommand{\arepo}{{\sc arepo}}          % code names
   \newcommand{\pluto}{{\sc pluto}}          % code names
   \newcommand{\flash}{{\sc flash}}          % code names
   \newcommand{\Fig}[1]{Fig.~\ref{fig:#1}}    % Fig. reference
   \newcommand{\Figure}[1]{Figure~\ref{fig:#1}}    % Figure reference

   \newcommand{\Aone}{{\sf A1}}
   \newcommand{\Atwo}{{\sf A2}}
   \newcommand{\Athree}{{\sf A3}}
   \newcommand{\Ione}{{\sf I1}}
   \newcommand{\Itwo}{{\sf I2}}
   \newcommand{\Ithree}{{\sf I3}}
   \newcommand{\Aonet}{{\sf A1\_t}}
   \newcommand{\Atwot}{{\sf A2\_t}}
   \newcommand{\Athreet}{{\sf A3\_t}}
   \newcommand{\Ionet}{{\sf I1\_t}}
   \newcommand{\Itwot}{{\sf I2\_t}}
   \newcommand{\Ithreet}{{\sf I3\_t}}
   \newcommand{\Aonetnw}{{\sf A1\_t\_nw}}
   \newcommand{\Atwotnw}{{\sf A2\_t\_nw}}
   \newcommand{\Athreetnw}{{\sf A3\_t\_nw}}
   \newcommand{\Ionetnw}{{\sf I1\_t\_nw}}
   \newcommand{\Itwotnw}{{\sf I2\_t\_nw}}
   \newcommand{\Ithreetnw}{{\sf I3\_t\_nw}}

   \newcommand{\IrrIonet}{{\sf IrrI1\_t}}
   \newcommand{\IrrItwot}{{\sf IrrI2\_t}}
   \newcommand{\IrrIthreet}{{\sf IrrI3\_t}}

   \newcommand{\CAone}{{\sf CA1}}
   \newcommand{\CAtwo}{{\sf CA2}}
   \newcommand{\CIone}{{\sf CI1}}
   \newcommand{\CItwo}{{\sf CI2}}
   \newcommand{\CAonet}{{\sf CA1\_t}}
   \newcommand{\CAtwot}{{\sf CA2\_t}}
   \newcommand{\CIonet}{{\sf CI1\_t}}
   \newcommand{\CItwot}{{\sf CI2\_t}}
   \newcommand{\CAonetnw}{{\sf CA1\_t\_nw}}
   \newcommand{\CAtwotnw}{{\sf CA2\_t\_nw}}
   \newcommand{\CIonetnw}{{\sf CI1\_t\_nw}}
   \newcommand{\CItwotnw}{{\sf CI2\_t\_nw}}

   \title{Late encounter-events as a source of disks and spiral structures}

   \subtitle{Forming second generation disks}

   \author{M. Kuffmeier
          \inst{1}\fnmsep\thanks{International Postdoctoral Fellow of Independent Research Fund Denmark (IRFD)}
          \and
          F. G. Goicovic\inst{1}
          \and
          C. P. Dullemond\inst{1}
          }

   \institute{Zentrum für Astronomie der Universität Heidelberg, Institut für Theoretische Astrophysik, Albert-Ueberle-Str. 2, 69120 Heidelberg
              \email{kuffmeier@uni-heidelberg.de}
             }

   \date{Received \today}

% \abstract{}{}{}{}{} 
% 5 {} token are mandatory
 
  \abstract
   {Both observations of arc-like structures and luminosity bursts of stars $> 1$ Myr in age indicate that at least some stars undergo late infall events. }
   {We investigate scenarios of replenishing the mass reservoir around a star via capturing and infalling events of cloudlets. }
   {We carry out altogether 24 three-dimensional hydrodynamical simulations of cloudlet encounters with a Herbig star of mass $2.5$ M$_{\odot}$ using the moving mesh code \arepo. To account for the two possibilities of a star or a cloudlet traveling through the interstellar medium (ISM), we put either the star or the cloudlet at rest with respect to the background gas.  }
   {For absent cooling in the adiabatic runs, almost none of the cloudlet gas is captured due to high thermal pressure. However, second-generation disks easily form when accounting for cooling of the gas. The disk radii range from several $100$ au to $\sim$1000 au and associated arc-like structures up to $10^4$ au in length form around the star for runs with and without stellar irradiation.
   Consistent with angular momentum conservation, the arcs and disks are larger for larger impact parameters.  
   Accounting for turbulence in the cloudlet only mildly changes the model outcome. 
   In the case of the star being at rest with the background gas, the disk formation and mass replenishment process is more pronounced and the associated arc-shaped streamers are longer-lived.} 
   {  The results of our models confirm that late encounter events lead to the formation of transitional disks associated with arc-shaped structures such as observed for AB Aurigae or HD 100546. In addition, we find that second-generation disks and their associated filamentary arms are longer lived ($>10^5$ yrs) in infall events, when the star is at rest with the background gas.
    }


   \maketitle
%
%________________________________________________________________

\section{Introduction}
Stars form and evolve in the dynamically evolving medium of Giant Molecular Clouds \citep[GMC;][]{Blitz1993}. They form as a consequence of gravitational collapse of so-called prestellar cores, which themselves emerge as a consequence of the existing turbulence in the GMC \citep{Padoan_turbfrag,MacLowKlessen2004}. 
Protostars grow in mass as long as they are sufficiently fed with mass from the circumstellar envelope of gas and dust \citep{Larson1969}. 
Models based on the assumption of a collapsing spherical prestellar core in isolation show that this phase typically lasts for $\sim100$ kyr \citep[e.g.][]{Machida2010,Masson2016}. 
During this initial time interval, a disk forms around the protostar \citep[e.g.][]{Machida2011,Li2011,Seifried2013,Tomida2015,Wurster2018}. 
These disks are the birthplaces of planets \citep{Keppler2018} and therefore commonly referred to as protoplanetary disks. 
Their lifetimes range between $\sim1$ and several Myr \citep{Haisch2001,Mamajek2009} before they dissipate. 
However, the formation process of stars is heterogeneous depending on the environment in which the protostar is embedded \citep{Kuffmeier2017,Kuffmeier2018}. 
Previous models already demonstrated the possibility of late accretion events around protostars \citep{Baraffe2009,Vorobyov2010,Padoan2014} as a possible explanation of the observed spread in protostellar luminosities \citep{Kenyon1990,Evans2009,Dunham2010}. 
In fact, gaseous condensations of size $\sim100$ au with masses $\ll1$ M$_{\odot}$ referred to as `cloudlets' have been observed previously \citep{Langer1995,Heiles1997,Falgarone2004}. 

Apart from that, late accretion may not only directly affect the properties of the accreting star \citep{Kunitomo2017,Jensen2018}, 
but also the properties of its surrounding \citep{Jorgensen2015,Frimann2017}, in particular its protoplanetary disk. 
The possibility of accretion onto star-disk systems has been studied in previous works by \citet{MoeckelThroop2009,Wijnen2016,Wijnen2017a,Wijnen2017b}.
Against the background of planet formation and the existence of disks, replenishing the mass reservoir of disks via late accretion is one possible solution to the `missing mass problem' in planet formation theory \citep{Mulders2015,Kuffmeier2017,Nixon2018,Manara2018}.   
The possibility of late mass accrual after dispersal of the primordial disk also raises the following question: if new material can fall onto a star hundreds of thousand of years after its formation, would it be possible to form a new disk again? 
\citet{Dullemond2019} (hereafter referred to as D19) considered the scenario, in which a star that is traveling through the interstellar medium captures gas from a condensation of gas in the Giant Molecular Cloud (GMC) after the priomordial disk has already dissipated.    
The scenario is referred to as `cloudlet capture' and it is related to the Bondi-Hoyle-Lyttleton (hereafter BHL) process \citep{HoyleLyttleton1939,BondiHoyle1944,Bondi1952}. Cloudlet capture differs from BHL accretion in the sense that the BHL process considers direct accretion onto the star, while in the cloudlet scenario, we investigate the effect of angular momentum of the cloudlet on the morphology of the gas around the moving star. In particular, we are interested, whether cloudlet capture events can lead to the formation of a new generation of disks and/or to the formation of arcs and `streamers' such as observed for e.g. AB Aurigae \citep{NakajimaGolimowski1995,Grady1999,Fukagawa2004}, FU Ori, Z CMa or V1057Cyg \citep{Liu2016,Liu2018}.  

Apart from the scenario of cloudlet capture, where the star travels through the interstellar medium and potentially captures gas from its surrounding, one can also consider a similar scenario of a cloudlet that actively moves toward a star at rest with its surrounding. We refer to this scenario as infall scenario. 
While being mostly equal by virtue of Galilei invariance, the difference of the two scenarios is the velocity of the low-density `space-filling' background medium in which both the star and the cloudlet are embedded.
In this paper, we consider both scenarios of cloudlet capture and infalling cloudlets.
We carry out altogether 24 models in a parameter study with the moving-mesh code \arepo, to investigate the consequences of captured and infalling cloudlets at a stage after a star has already formed in its parental GMC.   

We structure the paper as follows: in section 2, we present the different model setups. In section 3, we present the results of our parameter study. In particular, we present results from models with increasing complexity by first considering the setup of D19 as a default and consistently varying the setup by adding turbulence, removing the background velocity and accounting for stellar irradiation. 
We compare our results with observed structures and also discuss the limitations of our model in section 4. In section 5, we then summarize our main results and provide the main conclusions.

%__________________________________________________________________

\section{Methods}
\begin{table*}[th]
\centering
\begin{tabular}{llllllllll}
      & EOS & $R_{\rm cloudlet}$ {[}au{]} & $b$ {[}au{]} & $v_{\rm inf}$ {[}cm s$^{-1}${]} & $T_{\rm cloudlet}$ {[}K{]} & $M_*/M_{\odot}$ & $b_{\rm crit}$ & turb? & $v_{\rm background}$ \\ \hline
\Aone         & adia              &     887 & 1774 & 1.0 & 30 & 2.5 & 2218 & no  & yes\\
\Atwo         & adia              &    1330 & 1774 & 1.0 & 30 & 2.5 & 2218 & no  & yes\\
\Athree       & adia              &    2662 & 2218 & 1.0 & 30 & 2.5 & 2218 & no  & yes\\
\Ione         & iso               &     887 & 1774 & 1.0 & 10 & 2.5 & 2218 & no  & yes\\
\Itwo         & iso               &    1330 & 1774 & 1.0 & 10 & 2.5 & 2218 & no  & yes\\
\Ithree       & iso               &    2662 & 2218 & 1.0 & 10 & 2.5 & 2218 & no  & yes\\ \hline
\Aonet        & adia              &     887 & 1774 & 1.0 & 30 & 2.5 & 2218 & yes & yes\\
\Atwot        & adia              &    1330 & 1774 & 1.0 & 30 & 2.5 & 2218 & yes & yes\\
\Athreet      & adia              &    2662 & 2218 & 1.0 & 30 & 2.5 & 2218 & yes & yes\\
\Ionet        & iso               &     887 & 1774 & 1.0 & 10 & 2.5 & 2218 & yes & yes\\
\Itwot        & iso               &    1330 & 1774 & 1.0 & 10 & 2.5 & 2218 & yes & yes\\
\Ithreet      & iso               &    2662 & 2218 & 1.0 & 10 & 2.5 & 2218 & yes & yes\\ \hline
\Aonetnw      & adia              &     887 & 1774 & 1.0 & 30 & 2.5 & 2218 & yes & no \\
\Atwotnw      & adia              &    1330 & 1774 & 1.0 & 30 & 2.5 & 2218 & yes & no \\
\Athreetnw    & adia              &    2662 & 2218 & 1.0 & 30 & 2.5 & 2218 & yes & no \\
\Ionetnw      & iso               &     887 & 1774 & 1.0 & 10 & 2.5 & 2218 & yes & no \\
\Itwotnw      & iso               &    1330 & 1774 & 1.0 & 10 & 2.5 & 2218 & yes & no \\
\Ithreetnw    & iso               &    2662 & 2218 & 1.0 & 10 & 2.5 & 2218 & yes & no \\ \hline
\IrrIonet     & $T_{\rm irr}(r)$  &     887 & 1774 & 1.0 & 10 & 2.5 & 2218 & yes & yes\\
\IrrItwot     & $T_{\rm irr}(r)$  &    1330 & 1774 & 1.0 & 10 & 2.5 & 2218 & yes & yes\\
\IrrIthreet   & $T_{\rm irr}(r)$  &    2662 & 2218 & 1.0 & 10 & 2.5 & 2218 & yes & yes\\ 
\IrrIonet0    & $T_{\rm irr}(r)$  &     887 & 0    & 1.0 & 10 & 2.5 & 2218 & yes & yes\\
\IrrIonet500  & $T_{\rm irr}(r)$  &     887 & 500  & 1.0 & 10 & 2.5 & 2218 & yes & yes\\
\IrrIonet1000 & $T_{\rm irr}(r)$  &     887 & 1000 & 1.0 & 10 & 2.5 & 2218 & yes & yes\\

\end{tabular}
\caption{Summary of the setups of the different runs. First column: label of the individual run; second column: thermal prescription; 'adia' stands for adiabatic, 'iso' for isothermal and $T_{\rm irr}(r)$ for stellar irradiation. Third column: initial cloudlet radius $R_{\rm cloudlet}$ in au; fourth column: impact parameter $b$ in au; fifth column: initial bulk velocity of the cloudlet $v_{\rm inf}$ in cm s$^{-1}$; sixth column: initial temperature of the cloudlet $T_{\rm cloudlet}$ in K; seventh column: stellar mass in solar mass M$_{\odot}$; eighth column: critical impact parameter $b_{\rm crit}$ for which the deflection angle is $\pi/2$; ninth column: 'yes' and 'no' indicate, whether the turbulence description is used for the initialization of the cloudlet; tenth column: 'yes' and 'no' indicate, whether the background gas has the same initial velocity as the cloudlet ('yes') or whether it is at rest with respect to the star ('no').}
\label{runover}
\end{table*}

We base our 3D models of cloudlets being captured by a star on the simulations presented in D19, which modeled the interaction using the \pluto\ code \citep{Mignone2007}.
However, for this work we are simulating the encounter using the moving-mesh hydrodynamics code \arepo\ \citep{Springel2010,Pakmor2016}.
This code solves the Euler equations using a finite-volume approach on an unstructured Voronoi mesh that move with the flow.
As explained by \citet{Goicovic2019} in the context of stars being disrupted by supermassive black holes, \arepo\ is well suited for this kind of problems due to its quasi-Lagrangian nature, which retains the high accuracy of mesh-based techniques, while not imposing any preferred grid orientation.
Additionally, the fluxes are computed in the reference frame of the mesh faces, which greatly reduces advection errors, and does not lose accuracy with increasing gas velocity.
Finally, the full spacial and temporal adaptivity of \arepo\ allows us to refine or derefine cells as desired, similar to adaptive mesh refinement codes. In the case of our models, this allows us to maintain constant mass resolution in the cloudlet, while keeping a minimum spatial resolution in the low density regions of the ambient medium.

Based on the description by \cite{KlessenHennebelle2010}, we use as in D19 the following equation for the cloudlet mass 
\begin{equation}
    M_{\rm cloudlet}(R_{\rm cloudlet}) = 0.01 {\rm M}_{\odot} \left( \frac{R_{\rm cloudlet}}{5000 \rm au}\right)^{2.3},
    \label{KleHen}
\end{equation}
such that the mass solely depends on the radius of the cloudlet $R_{\rm cloudlet}$.
The star is modeled as an external Newtonian potential, located at the center of the domain with an initial mass of $2.5$ M$_{\odot}$.
In order to avoid divergent accelerations, this gravitational potential has a softening radius of 5 au around the center.
Furthermore, since this scale is significantly smaller than the typical size of the gas cells, in the vicinity of the star we impose an additional refinement criterion to the ones already mentioned to maintain a certain number of gas cells per softening length (see Appendix~\ref{app:refinement}).
Finally, to mimic the capture of gas by the star, and to reduce the computational cost of these models, we implement accretion onto the central potential. Within 25 au of the center, all gas cells are drained of 90\% of their mass at each timestep and directly added to the central mass.
While the star is at rest, the cloudlet and the surrounding gas has an initial bulk velocity of $v=(v_{\infty},0,0)$, where $v_{\infty}$ is set to $10^5 \unit{cm}\unit{s}^{-1}$.

We model the cloudlet capture in a large box with periodic boundary conditions.
The box length is defined as $L_{\rm box} = 148 R_{\rm cloudlet}$. We stop the simulations early enough that the results are unaffected by the assumption of periodic boundary conditions.
Altogether, we conduct a series of 24 models as summarized in Table \ref{runover}.
The basic setups are the same as considered in D19 for instantaneous cooling (isothermal cases \Ione, \Itwo, \Ithree) and without cooling (adiabatic cases \Aone, \Atwo, \Athree). In other words, we use a ratio of specific heat of $\gamma=\frac{5}{3}$ in the adiabatic case, and $\gamma=1$ for the isothermal runs.
In the adiabatic runs, we insert the center of the cloud at $(x,y,z)=(-7.82 R_{\rm cloudlet},-b,0)$ with $b$ being the impact parameter. 
We set the background temperature to $T_{\rm bg}=8000$ K inspired by the temperature of a warm neutral medium \citep{Field1969}. The cloudlet temperature of the adiabatic runs is $T_{\rm cloudlet}=30$ K, and the background density is $\rho_{\rm bg} = \rho_{\rm cloudlet} \frac{T_{\rm cloudlet}}{T_{\rm bg}}$. 
This choice guarantees that the cloudlet is pressure confined in the adiabatic runs. 

For the isothermal runs we set the temperature to $T=T_{\rm cloudlet}=T_{\rm bg}=10$ K everywhere in the domain, and we set the background density to $\rho_{\rm bg} = \rho_{\rm cloudlet} \frac{T}{8000 \unit{K}}$.
As the cloudlet is not pressure confined in the isothermal run, it immediately starts to expand. 
To nevertheless model the capturing and infalling process, we place the cloudlet initially closer to the central potential at a location of $(x,y,z)=(-3.22 R_{\rm cloudlet},-b,0)$. 

Additionally, we carry out a set of models in which we add turbulence to the initial cloud setups. These runs are labeled with an additional `\textsf{\_t}'.
The turbulent velocity of each gas cell is drawn from Gaussian random distribution using the prescription described in \citet{Dubinski1995}. 
We sample the velocities in Fourier space with a a power spectrum of $|\vb{v}_k|^2\propto k^{-4}$, where $k$ is the wavenumber of the perturbation, and the power index is chosen to match the velocity dispersion observed in molecular clouds \citep{Larson1981}.
Finally, the velocity field is normalized such that the internal velocity dispersion of the cloudlet is 10\% of its initial bulk speed, i.e. we multiply each velocity by
\begin{equation}
    A_v = 0.1 \frac{v_\infty}{\sigma_v},
\end{equation}
where $\sigma_v$ is the velocity dispersion of field before normalization.
Notice that, with this choice of parameters, the internal turbulence is supersonic with typical Mach numbers around 4.

\begin{figure*}
  \centering
  \includegraphics[trim=200 208 0 304,clip, width=\textwidth]{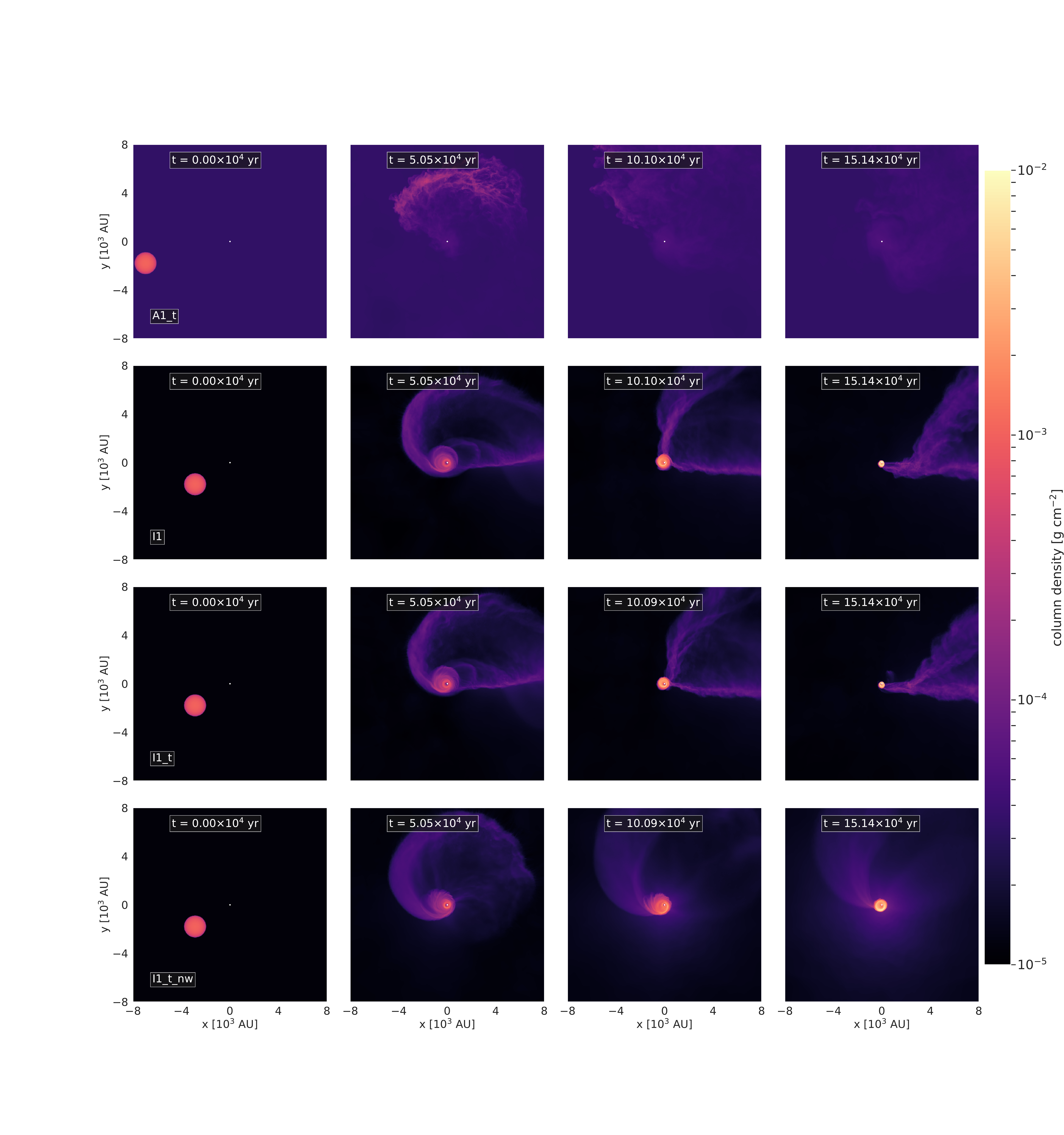}
  \caption{Illustration of the column density in the planes perpendicular to the orientation of the angular momentum vector for the setups of the adiabatic run \Aone\ (upper panels) and the isothermal run \Ione\ (lower panels) at times ($t_0=0$, $t_1=4 t_{\rm dyn}$, $t_2=8 t_{\rm dyn}$ and $t_3 =16 t_{\rm dyn}$.}
  \label{fig:prj_A1I1}
\end{figure*}

The models presented up to this point consider the scenario where the star travels through the medium at a given speed and captures mass from its surroundings,
but we are also interested in the related scenario of infall. Here, only the individual gas condensation has a relative velocity difference compared to the star co-moving with the background velocity. We model this scenario by only giving a velocity of $v= (v_{\infty},0,0)$ to the cloudlet, while setting the background gas and the central mass potential at rest. The infall models are labeled with an additional \textsf{\_nw} to indicate that the background gas is at rest (`no wind' of the background gas). 

Furthermore, in order to more adequately model the presence of the star beyond its gravitational influence, it is important to consider that stellar irradiation heats up the surrounding gas.
To mimic this effect, we adopt a radial temperature profile that considers the irradiation of a perfect black body, constant in time, as follows
\begin{equation}
    T(r) = \mathrm{max} \left( 10 \unit{K}, \left( \frac{L_{*} }{16\pi \sigma_{\rm SB} r^2 } \right)^{1/4} \right),
    \label{eq:T_r}
\end{equation}
where $L_*$ is the luminosity of the star, $\sigma_{\rm SB}$ is the Stefan-Boltzmann constant and $r$ is the radial distance from the star, i.e., the center of our domain.
Applying this recipe, the temperature is 10 K at radial distances of $r>776 \unit{au}$ from the star. We refer to the runs accounting for stellar irradiation with \textsf{Irr}, and we investigate the effect of this heating implementation using the setups of the isothermal runs.  

Additionally, using the stellar irradiation recipe, we carry out three runs with varying impact parameter $b$, which we label by appending the value of $b$ in au, namely \textsf{0}, \textsf{500} and \textsf{1000}. 
Except for $b$, the initial size and location of the cloudlet is identical to the setup of \IrrIonet.

The cloudlet is initially resolved with a total of $10^6$ cells and we smoothly decrease the resolution beyond the cloudlet to a lower background resolution as illustrated in \Fig{grid_setup}. Starting with this smooth transition in the initial setup prevents spurious shock waves from being launched at the edge of the cloud immediately after starting the simulations. 
All models are run without self-gravity, hence the central mass is the only source of gravity.

\section{Results}
We first give a qualitative description of the sequence of cloudlet capturing and infall for the adiabatic and isothermal setups. In the following, we show results from runs in which we account for stellar irradiation, and quantify the encounter events by focusing on the amount of captured mass as well as on the properties of forming disks. 

\subsection{Sequence for the adiabatic and isothermal setups}

\begin{figure*}
  \centering
  \includegraphics[trim=50 228 10 304,clip, width=0.966\textwidth]{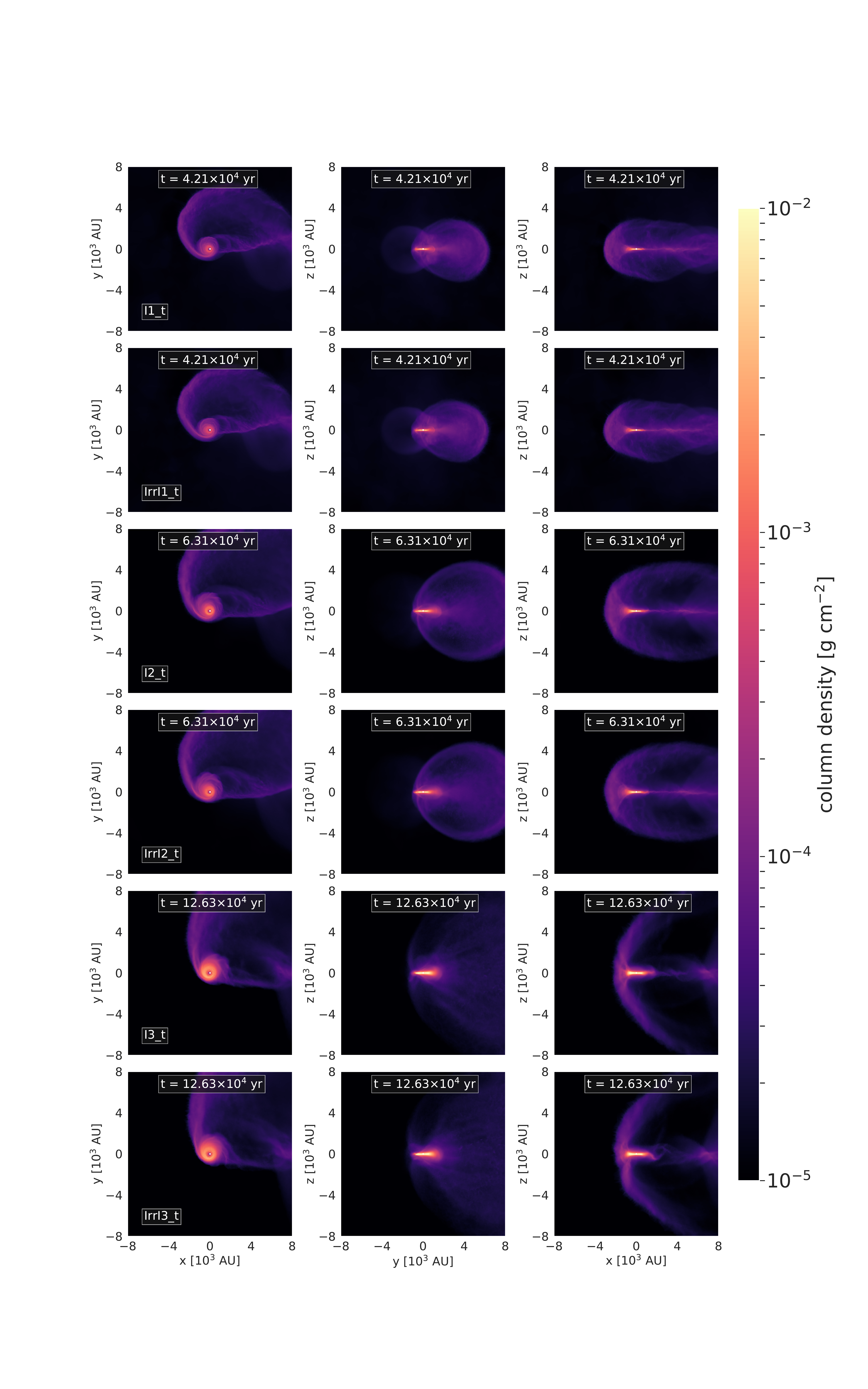}
  \caption{Illustration of the column density in the three planes of the coordinate system (left panels: xy, middle panels: yz, right panels: xz) for the setups of the isothermal runs after ten dynamical times ($t=10 t_{\rm dyn}$). Panels in odd rows correspond to pure isothermal runs, panels in even rows correspond to the cases with stellar irradiation.}
  \label{fig:prj_tencross}
\end{figure*}

\begin{figure*}
  \centering
  \includegraphics[trim=200 208 0 304,clip, width=\textwidth]{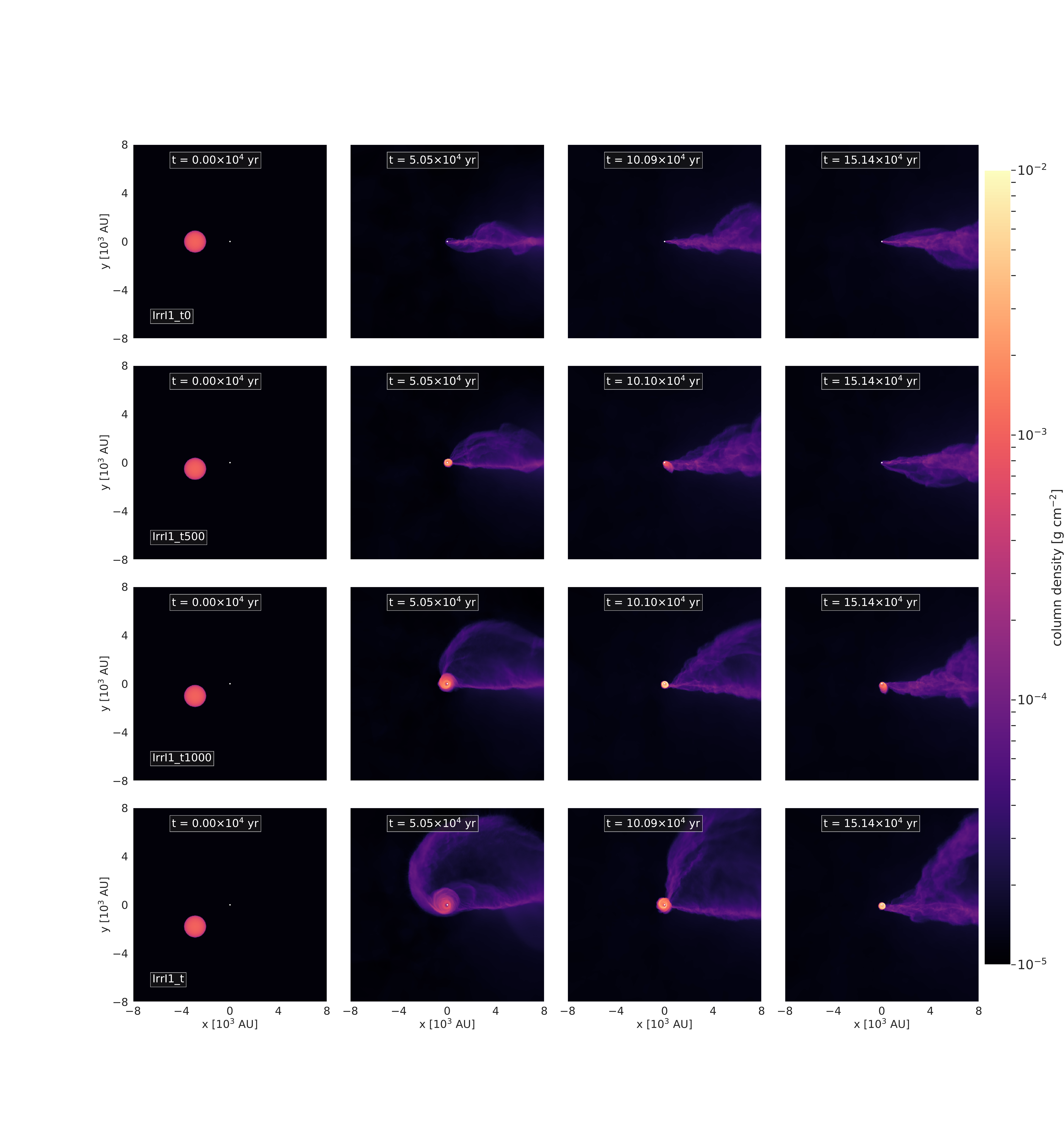}
  \caption{Column density in the planes perpendicular to the orientation of the angular momentum vector for runs \IrrIonet0, \IrrIonet500, \IrrIonet1000\ and \IrrIonet\ at times ($t_0=0$, $t_1=4 t_{\rm dyn}$, $t_2=8 t_{\rm dyn}$ and $t_3 =16 t_{\rm dyn}$.}
  \label{fig:prj_IrrI1}
\end{figure*}

\begin{figure*}
  \centering
  \includegraphics[width=\textwidth]{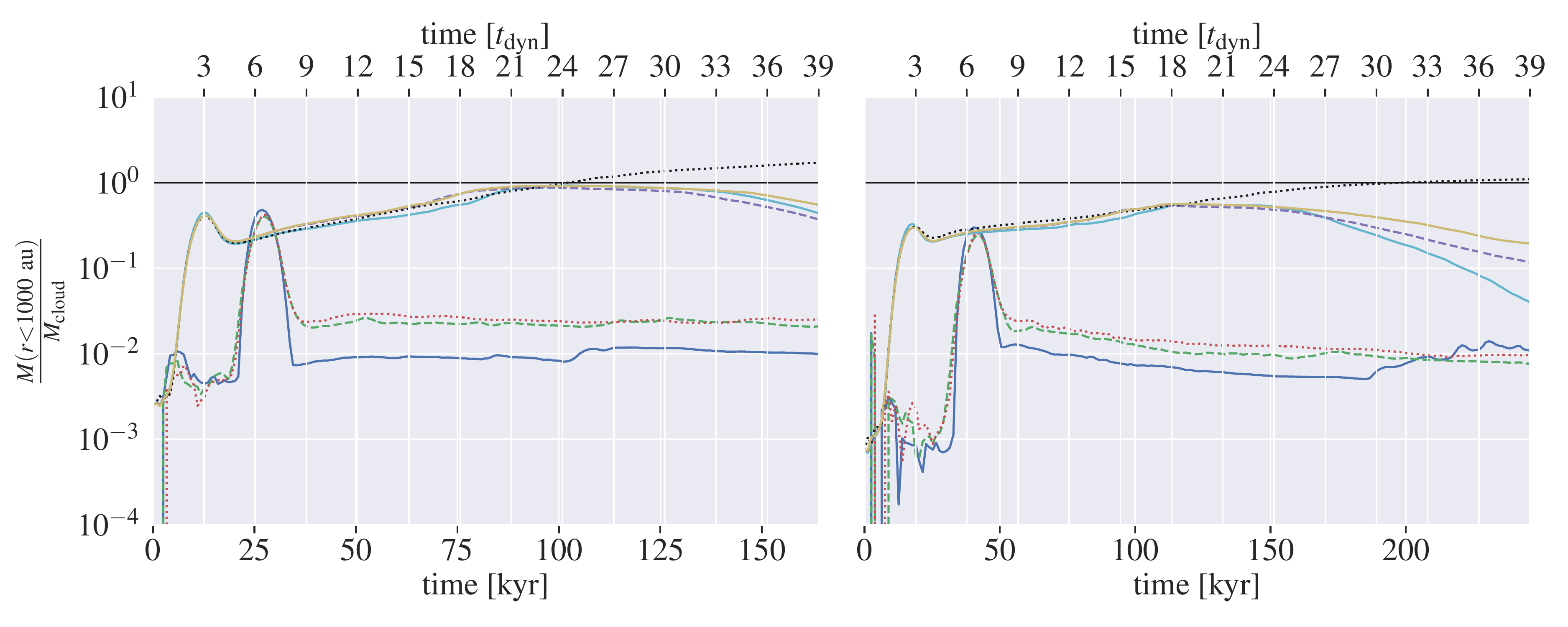} \\
  \includegraphics[width=\textwidth]{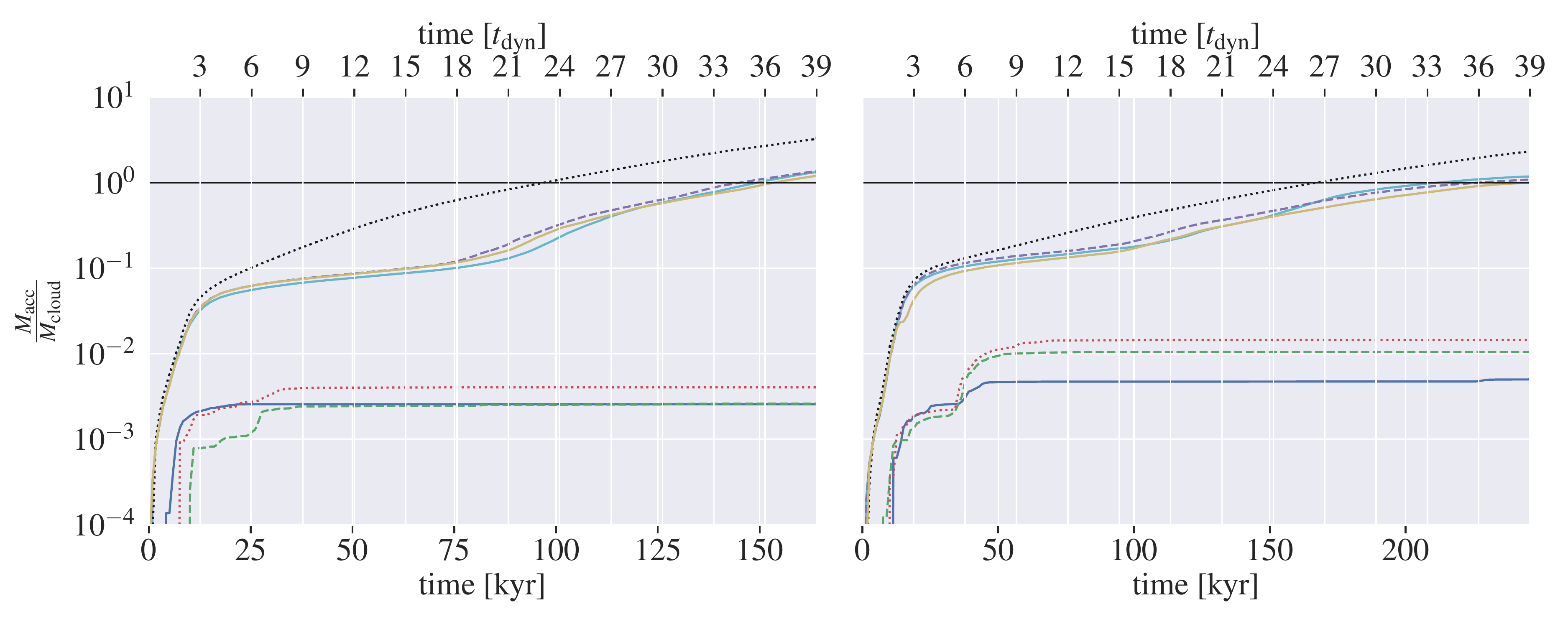}
  \caption{Evolution of captured mass and accreted mass for altogether 14 runs. The upper panels show the amount of mass within 1000 au from the central potential, and the lower panels display the corresponding amount of accreted mass. The line style and color is used consistently for the upper and lower plots. Left panels: \Aone\ (blue solid), \Aonet\ (green dashed), \Aonetnw\ (red dotted), \Ione\ (cyan solid), \Ionet (purple dashed), \Ionetnw\ (black dotted) and \IrrIonet\ (yellow solid), right panels: \Atwo\ (blue solid), \Atwot\ (green dashed), \Atwotnw\ (red dotted), \Itwo\ (cyan solid), \Itwot\ (purple dashed), \Itwotnw\ (black dotted) and \IrrItwot\ (yellow solid).}
  \label{fig:enclosed_mass}
\end{figure*}

In this subsection, we qualitatively present the process of encounter events using the thermodynamical description used in D19. 
In \Fig{prj_A1I1}, we illustrate column densities in the plane perpendicular to the angular momentum vector around the central star for the cases of \Aonet, \Ione, \Ionet\ and \Ionetnw\ at time $t=0, 12, 24, 36 t_{\rm dyn}$, where $t_{\rm dyn} = \frac{R_{\rm cloudlet}}{v_{\infty}}$ with $R_{\rm cloudlet}$ being the cloudlet radius and $v_{\infty}$ being the escape velocity, which we set to $1 \unit{km}\unit{s}^{-1} = 10^5 \unit{cm}\unit{s}^{-1}$. The projected area is $16000 \unit{au} \times 16000 \unit{au}$. The plots illustrate that gas is affected by the central mass in all setups. The process differs significantly between the adiabatic and isothermal models. In the adiabatic run, the gas from the cloudlet initially engulfs the star in a halo-like manner, and then forms a transient arc-like structure (as seen in the second panel). However, most of the gas passes by the star in a hyperbolic orbit due to its intrinsic angular momentum and disperses during the further evolution, such that almost no gas is captured in the adiabatic scenario (third and fourth panel). 

In the isothermal runs, we also see the formation of bent structures, but rather as a tail associated with the formation of disks than remnants of dispersing halos.  In contrast to the adiabatic scenario, mass is captured in the isothermal run, and disks of radii $r\gg100$ au easily form. The details of the process differ between the setups, but in general gas passes by in the adiabatic runs, while it can remain in the vicinity of the star and lead to the formation of disks in the isothermal runs. 

This behavior is expected considering that the isothermal and adiabatic setups are two extreme scenarios of either perfect or absent cooling. In the adiabatic run, thermal pressure increases when the gas gets compressed during the capturing phase. Therefore, the fresh gas leads to the formation of an intermittent envelope, but not to the formation of a disk. In the isothermal run, infalling gas remains cold, and due to conservation of angular momentum disks quickly form.
Our results for the models without turbulence are in good agreement with the results based on simulations with \pluto\ presented in D19.

As seen in the third row of \Fig{prj_A1I1}, the sequence is qualitatively the same regardless of whether we start with or without an initially turbulent cloudlet. The infall scenarios ($\textsf{\_nw}$ runs), where only the cloudlet is moving and the background gas is at rest, are similar during the early evolutionary phase, but differ at later times. Comparing the morphology of the arc-like structures, we find that in the capturing scenario of a traveling star, the bent arms are dragged by the motion of the background gas, leading to morphological differences of the arm structures around the star (see fourth column in \Fig{prj_A1I1}). 
In general, the disk shrinks in radius over time and more mass is swept away by the background velocity for the capturing scenarios (\Ione\ and \Ionet).  
In the infall scenario, mass remains in the vicinity of the star for a longer time correlating with longer lasting disks. 

\subsection{Toward more realistic modeling: accounting for stellar irradiation}
In the models presented above, we consider two extreme scenarios of perfect or absent cooling. 
In reality, dense gas in the ISM is cold, but the central star acts as a radiation source heating up its surrounding. We account for the effect of stellar irradiation by adopting the radially dependent temperature profile that is constant in time given by eq.\ \ref{eq:T_r}. For the luminosity of the star, we assume $L_* = 50 L_{\odot}$, similar to the luminosity of AB Aurigae with $L=47 L_{\odot}$ \citep{Tannirkulam2008}. We run these models for an initially turbulent cloudlet and with background velocity. Similar to the isothermal models, we find the formation of large disks after the encounter phase of the cloudlet with the star. 
As illustrated in the projection plots along the three different coordinate axes in \Fig{prj_tencross}, the results are only mildly different from the pure isothermal runs. This is not surprising considering the large radial extent of the disks far beyond $r=100$ au and considering that stellar irradiation only mildly increases the temperature of the initially cold gas ($T=10$ K) at large radii (e.g. $T(100 \unit{au})\approx74$ K, $T(1000 \unit{au})\approx23.5$ K). Although this temperature profile still is rather simplified, the models nevertheless demonstrate the possibility of newly forming disks for cloudlet encounters for already formed stars when accounting for stellar irradiation. In fact, we argue that we rather over- than underestimate heating by the central star because we neglect any shielding by the gas.
Moreover, the column density profiles in the left panels of \Fig{prj_tencross} illustrate the formation of large arc-structures extending to distances beyond several $10^3$ au from the star associated with second generation disks of several 100 au to $r\sim$1000 au in size.

\subsection{Varying the impact parameter $b$} 
As seen in D19 and confirmed in this work, the capturing process varies for different initial cloud parameters $R_{\rm cloudlet}$ and impact parameter $b$. To better constrain the effect of the impact parameter $b$, we varied $b$ for otherwise identical conditions as in \IrrIonet\ (runs \IrrIonet0, \IrrIonet500 and \IrrIonet1000). 
Analogous to \Fig{prj_A1I1}, we illustrate the evolution by showing snapshots at $t=0, 12, 24, 36 t_{\rm dyn}$ of the runs \IrrIonet0, \IrrIonet500, \IrrIonet1000\ and \IrrIonet\ in \Fig{prj_IrrI1}. In the case of $b=0$, i.e., zero net angular momentum of the cloudlet, we see that the cloudlet passes the central mass and gets disrupted without leaving any trace of forming a disk. Applying a non-zero impact parameter leads to the formation of disks with larger and longer lasting disks for increasing impact parameters, as expected from angular momentum conservation.
Moreover, in the cases of larger impact parameters $b=1000$ au and $b=1774$ au, associated with the formation of disks, we find the formation of arcs that are several 1000 au in length.  

Although the amount of enclosed mass and the properties of the disks quantitatively differ for the runs with different initial cloudlet size or impact parameter, the sequence is generally similar. 
Given a non-zero impact parameter, disks and bent arm structures form for models with cooling of the gas (isothermal and stellar irradiation runs), while disk formation is suppressed in the adiabatic runs.   

\subsection{Captured mass}
To quantify the capturing process more, we show the evolution of mass that is enclosed within 1000 au from the central star (upper panels) as well as the amount of accreted mass (lower panels) over 39 dynamical times $t_{\rm dyn}$ for the different runs with the setups of \Aone\ and \Ione\ (left panels), \Atwo\ and \Itwo\ (right panels) in \Fig{enclosed_mass}. We compute the amount of accreted as the cumulative sum of the difference of total mass in the simulation between the different snapshots. Only for a few snapshots after a simulation restart during a run, we are forced to interpolate the mass difference between the snapshots. In both the adiabatic and the isothermal setups, we can see a characteristic early peak in enclosed mass that is almost independent of whether the model is run with or without turbulence or background velocity. The peak occurs earlier in the isothermal setups as a consequence of initially placing the cloudlet closer to the star to prevent cloudlet dispersal before the encounter. For the smaller cloudlet radius (upper left panel), a larger fraction of cloudlet mass reenters the region around the star within $1000$ au after the initial fly-by, whereas relatively more cloudlet mass is swept away with the background velocity instead of being captured for the larger cloudlet radius (upper right panel).
For the setup with the largest cloudlet radius (\Ithree\ setups, not shown), the peak is almost entirely absent.  

Comparing the runs with identical impact parameter $b$ and cloudlet radius $R_{\rm cloudlet}$ among each other, the evolution differs significantly after the encounter of the cloudlet with the star.  
The plots demonstrate the fundamental difference of absent capturing in the adiabatic case and mass growth in the isothermal case. While in the isothermal runs, a substantial amount of mass is kept within a radial distance of 1000 au from the central star, the enclosed mass rapidly decreases again after the early peak during the encounter in the adiabatic run. In the adiabatic cases, the mass only marginally grows to about $\sim 10^{-2} M_{\rm cloudlet}$ with respect to the initial setup.

For the isothermal setups we find that the amount of enclosed mass decreases after the early peak, too. However, the amount of enclosed mass decays less after the initial peak than in the adiabatic runs. In sharp contrast to the adiabatic runs, the amount of captured mass increases again after the early enhancement. 
Especially, for relatively small cloudlet sizes (\Ione, \Itwo), we find significant mass growth at later evolutionary stages for the isothermal setups. The smaller the cloudlet radius $R_{\rm cloudlet}$, the more mass relative to the mass of the cloudlet is captured again after the peak during the encounter. Our analysis shows that the amount of enclosed mass exceeds the first peak for small enough $R_{\rm cloudlet}$.
The reason for this is twofold. First, given an identical impact parameter (such as for the setups 1 and 2), part of the gas is further away from the star for large cloudlets, and hence this gas is less affected by the gravitational potential of the star. Second, the density of the cloudlet is 
\begin{equation}
\rho_{\rm cloudlet} = \frac{M_{\rm cloudlet}}{4/3 \pi R_{\rm cloudlet}^3},
\end{equation}
and using eq. (\ref{KleHen}), we obtain a scaling of the density with initial cloudlet radius $R_{\rm cloudlet}$ as  
\begin{equation}
\rho_{\rm cloudlet} = R_{\rm cloudlet}^{-0.7}.
\end{equation}
Recalling that by construction the background density is $\rho_{\rm bg}=\frac{1}{800} \rho_{\rm cloudlet}$ shows that both the cloudlet as well as the background volume density are larger for smaller cloudlet sizes. 
As a consequence, more mass is available from the background for the models with smaller cloudlet radii, while the cloudlet has lower mass than for models with larger cloudlet radii. Additionally, smaller cloudlets disperse more quickly for the isothermal setups due to the larger density, and hence the larger pressure gradient between background gas and cloudlet gas.  
As seen for the scenarios of \Itwo\ and even more for the \Ione\ scenarios in \Fig{enclosed_mass}, this can even lead to a state, where the amount of enclosed mass is larger than the initial cloudlet size.   
The enclosed mass profiles show that for small enough impact parameters, part of the cloudlet mass flies by the star, departs from it, but bounces back due to the stellar gravitational potential, and hence gas can accrete onto the star in a trajectory as illustrated in the projection plots in \Fig{prj_A1I1}.

Concerning the differences among the models with initially identical $R_{\rm cloudlet}$ and $b$, we confirm that the amount of captured mass (upper panel in \Fig{enclosed_mass}) and accreted mass (lower panel in \Fig{enclosed_mass}) is almost identical, regardless whether the models are run with or without turbulence and with or without our recipe of stellar irradiation. 
Therefore, the results show that the velocity deviations due to turbulence in the cloudlet are of secondary importance for the overall process of mass replenishment around the star. 
However, during the later evolution we find stronger deviations among the profiles for the runs with background velocity compared to the infall scenario without background velocity. 
For the runs including the background velocity, we find a second peak in the mass profile, which is absent for the infall scenario without background velocity. 
For the runs with background velocity, the second enhancements in the mass profiles occur after less dynamical times for higher cloud radii. However, we point out that the concept of a dynamical time based on the initial cloudlet radius loses its meaning at later stages, when the cloudlet is in fact destroyed due to the combination of dispersal and the earlier encounter event.   

In the infall cases, more mass continuously falls toward the central star and replenishes the mass reservoir around the star (upper panel in \Fig{enclosed_mass}) and subsequently accretes onto the star (lower panel in \Fig{enclosed_mass}). As a consequence of the absent background wind that sweeps mass away from the central gravitational potential, the amount of accreted mass is significantly higher for the infall scenario compared to the other scenarios. 
Compared to the purely isothermal runs, we find a mildly higher amount of enclosed mass, but smaller amount of accreted mass at late times for the runs incorporating our stellar irradiation recipe (see in particular the results for setup 2 in the right panels of \Fig{enclosed_mass}. This difference is caused by the additional thermal pressure close to the star that prevents gas from accreting onto the central potential as directly as in the isothermal runs.

As expected from angular momentum conservation, we confirm that for identical cloudlet size $R_{\rm cloudlet}$, the smaller $b$, the quicker the encountering cloudlet mass accretes onto the central star. This is in line with the presence of larger disk and arc structures as described in the comparison of impact parameter $b$ (see \Fig{prj_IrrI1}). In the case of smaller $b$, a larger amount of mass is quickly 'lost' to the central star, and hence cannot lead to the formation of substantial secondary disks or arcs.  
In general, our results show that encounters with a cloudlet can efficiently enrich the mass reservoir around the star regardless of the details of the specific setups.

\begin{figure*}
  \centering
  \includegraphics[width=\textwidth]{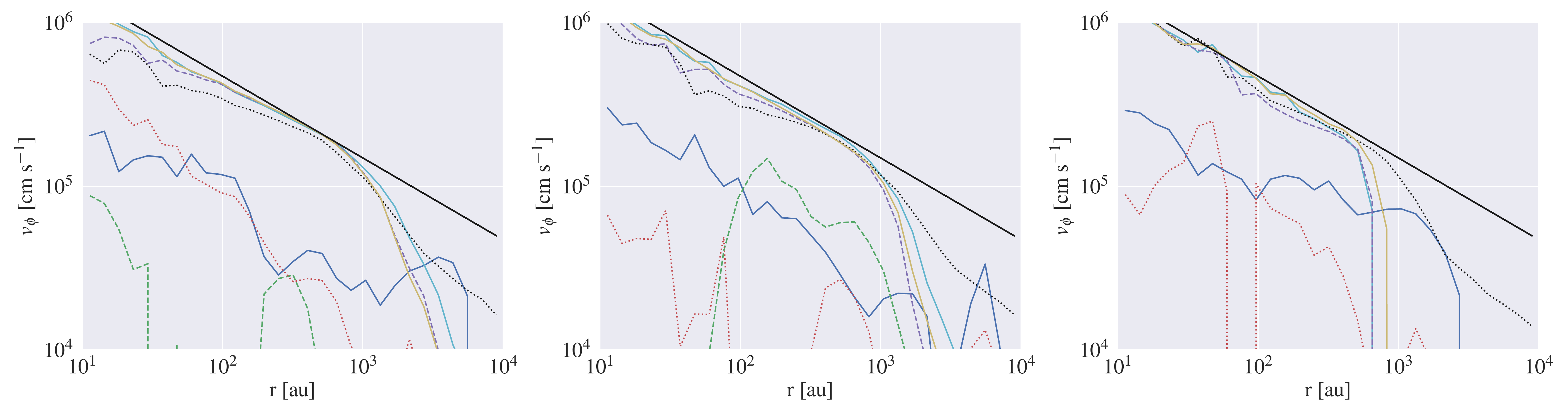}
  \caption{Velocity profiles at $t=16 t_{\rm dyn}$ for the different runs. 
  Left panel: \Aone\ (blue solid), \Aonet\ (green dashed), \Aonetnw\ (red dotted), \Ione\ (cyan solid), \Ionet (purple dashed), \Ionetnw\ (black dotted) and \IrrIonet\ (yellow solid), middle panel: \Atwo\ (blue solid), \Atwot\ (green dashed), \Atwotnw\ (red dotted), \Itwo\ (cyan solid), \Itwot\ (purple dashed), \Itwotnw\ (black dotted) and \IrrItwot\ (yellow solid), right panel: \Athree\ (blue solid), \Athreet\ (green dashed), \Athreetnw\ (red dotted), \Ithree\ (cyan solid), \Ithreet (purple dashed), \Ithreetnw\ (black dotted) and \IrrIthreet\ (yellow solid). The black solid line corresponds to the Keplerian velocity $v_{\rm K} = \sqrt{\frac{GM}{r}}$ for a star with mass $M=2.5 \unit{M}_{\odot}$.}
  \label{fig:vphi_16tdyn}
\end{figure*}

\begin{figure*}
  \centering
  \includegraphics[trim=170 20 0 80,clip,
  width=\textwidth]{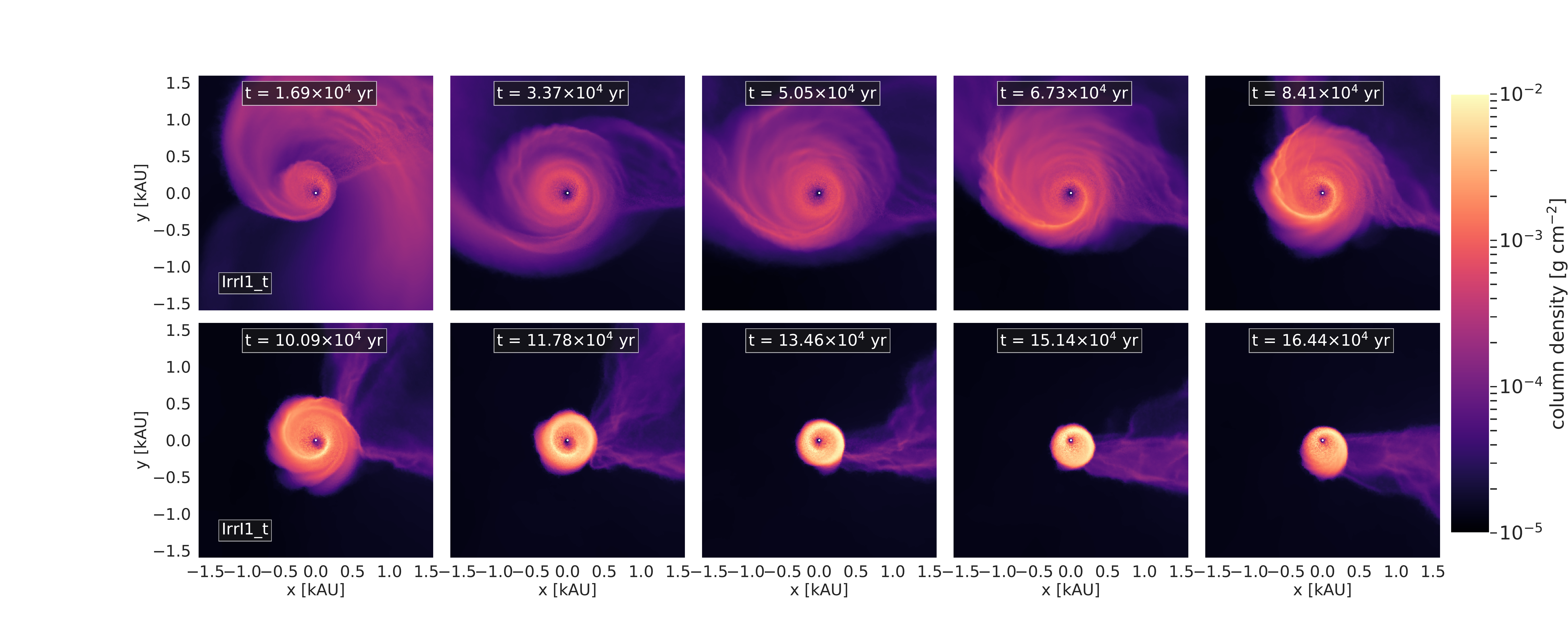}
  \caption{Column density in the planes perpendicular to the orientation of the angular momentum vector for run \IrrIonet\ at different times. The upper panel displays from left to right $t_1=4 t_{\rm dyn}$, $t_2=8 t_{\rm dyn}$, $t_3 =12 t_{\rm dyn}$, $t_4=16$, $t_5=20 t_{\rm dyn}$, and the lower panel displays from left to right $t_6=24 t_{\rm dyn}$, $t_7 =28 t_{\rm dyn}$, $t_8=32 t_{\rm dyn}$, $t_9=36 t_{\rm dyn}$ and at the end of the simulation $t_{10}=39 t_{\rm dyn}$. }
  \label{fig:prj_IrrI1_small}
\end{figure*}

\begin{figure}
  \centering
  \includegraphics[width=\columnwidth]{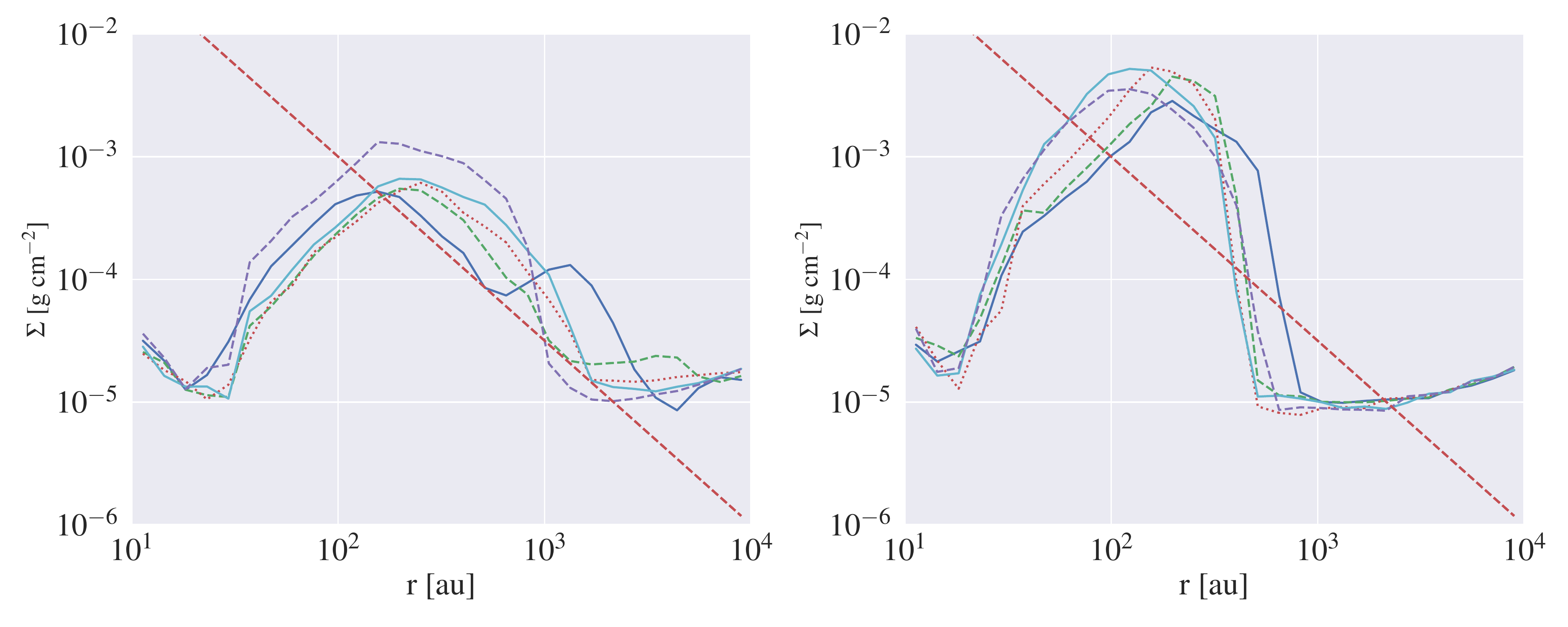}
  \caption{Azimuthally averaged column density in the planes. The column densities are computed perpendicular to the orientation of the angular momentum vector for run \IrrIonet. The profiles correspond to the times displayed in the projection plots in \Fig{prj_IrrI1_small}. Left panel: $t_1=4 t_{\rm dyn}$ (blue solid), $t_2=8 t_{\rm dyn}$ (green dashed), $t_3 =12 t_{\rm dyn}$ (red dotted), $t_4=16$ (cyan solid), $t_5=20 t_{\rm dyn}$ (purple dashed), right panel: $t_6=24 t_{\rm dyn}$ (blue solid), $t_7 =28 t_{\rm dyn}$ (green dashed), $t_8=32 t_{\rm dyn}$ (red dotted), $t_9=36 t_{\rm dyn}$ (cyan solid), $t_{10}=39 t_{\rm dyn}$ (purple dashed). The red dashed line corresponds to the profile of the MMSN, $\Sigma = 1700 \unit{g} \unit{cm}^{-2} \left(\frac{r}{\unit{au}}\right)^{-3/2}$.}
  \label{fig:Sigma_I1evo}
\end{figure}

\subsection{Second generation disk formation}

\subsubsection{Velocity profile}
As illustrated in the exemplary projection plots in \Fig{prj_A1I1}, \Fig{prj_tencross} and \Fig{prj_IrrI1}, disks form for the isothermal runs with and without stellar irradiation. 
To better constrain the properties of the velocity field, we compute the average velocity in altogether 30 shells with logarithmically increasing radius $r_i$ from 1 au to $10^4$ au around the star. The height of each bin is set to $h=\pm 0.1 r_i$ and the orientation is such that the radial direction is perpendicular to the orientation of the total angular momentum vector computed for the gas located within 1000 au from the star.   
In \Fig{vphi_16tdyn}, we show the velocity profile at $t=16 t_{\rm dyn}$ for the different models (without turbulence, with turbulence, infall with turbulence) of \Aone, \Ione, \Atwo, \Itwo, \Athree\ and \Ithree. The plot demonstrates that the velocity profiles of the disks in the isothermal and stellar irradiation runs are in good agreement with the profile of a Keplerian disks, while the adiabatic runs typically do not show any signs of Keplerian rotation (except for a mild signature in the \Athree\ run). The velocity profiles show that the isothermal disks typically extend to radii of $\approx500$ au to $\approx1000$ au.

Comparing the isothermal models among each other (without turbulence, with turbulence, with turbulence and background velocity) shows that the models only slightly deviate for radii smaller than $r\lesssim 1000$ au. 
As showed above, the most significant differences occur between the infall scenario without background velocity compared to the scenarios with background velocities. As a consequence of enhanced accretion onto the central sink, the azimuthal velocity in the infall runs of \Ione\ and \Itwo\ is smaller than for the runs with background velocity for $r\lesssim 1000$ au. Beyond that radius the velocity falls off more sharply in the runs with background velocity than in the runs without background velocity. 
This tendency can be seen more clearly for the runs with larger cloudlet radii (\Itwo\ in the middle panel and \Ithree\ in the right panel). The more shallow decay of the rotational velocity profile is induced by the fact that the background starts spiraling toward the central star, while in the other runs the velocity profiles of the lower density gas significantly deviates from Keplerian as a consequence of the intrinsic background flow.     
Regardless of the subtle differences among the different setups, the results of our isothermal and stellar irradiation runs show that the disks are indeed rotationally supported and relatively large as expected from angular momentum conservation.

\subsubsection{Column density profile}

The scope of this paper is to investigate the overall possibility, whether mass replenishment induced by cloudlet encounters leads to the formation of bent structures and second-generation disks, rather than carrying out a detailed study of the properties of the corresponding disks. 
However, the distribution of mass in the disk is a crucial parameter, when speculating about planet formation in a second-generation disk.
Therefore, we analyze the evolution of the disks in our \IrrIonet\ in more detail. To illustrate the sequence of disk formation and evolution more clearly, we show projection plots in the plane of the disk for a region of $3200 \unit{au} \times 3200 \unit{au}$ at ten different times in \Fig{prj_IrrI1_small}. Furthermore, we show the azimuthally averaged profiles of the surface density at the corresponding times in \Fig{Sigma_I1evo} (left panel: $4 t_{\rm dyn}$, $8 t_{\rm dyn}$, $12 t_{\rm dyn}$, $16 t_{\rm dyn}$ and $20 t_{\rm dyn}$, right panel: $24 t_{\rm dyn}$, $28 t_{\rm dyn}$, $32 t_{\rm dyn}$, $36 t_{\rm dyn}$ and $39 t_{\rm dyn}$). 
We use the same radial spacing as for the computation of the velocity profile and the columns extend to the entire size of the domain.  

Generally, we find that the surface density profiles undergo a typical sequence. During the encounter, a disk starts to form with maximum density at $r\approx 100$ au and connected to the background with a larger spiral arm. 
During the further evolution, the disk becomes more compact leading to an increase of $\Sigma$ in the range of a few $100$ au to $\approx$1000 au, and erosion of the outer part of the disk due to the background velocity. Over time, the bump, in $\Sigma$ decreases again and migrates toward smaller radii (see \Fig{Sigma_I1evo}). Interestingly, a spiral arm structure decoupled from the larger scale environment starts to form in the disk and compactifies together with the shrinking disk. 
The compactification phase leads to an enhancement of density at a radius of a few 100 au with a steep drop inside and outside the evolving bump. That  means  that  consistent  with  the  mass
distributions in transitional disks most of the gas is located in
the outer part of the disk though the inner decrease in density is favored by our accretion recipe.

Gas located in the vicinity of the central potential accretes and causes a drop in the surface density profile. Simultaneously, gas with initially large angular momentum contributes to the mass budget of the disk. The outer edge of the disk becomes steeper because the high-density gas initially located in the cloudlet has either already fallen onto the disk or is continuously swept away in the case of a present background wind. 

For comparison, in \Fig{Sigma_I1evo}, we also plot the slope of the minimum mass solar nebula (MMSN) $\Sigma \propto r^{-\frac{3}{2}}$ in the azimuthally averaged column density profile. During the early formation of the disk, the column density profile is similar to a power-law function in the range of $\sim$100 au to $\sim$1000 au that is roughly consistent with the slope of the MMSN. However, the radial extent is rather large compared to a primordial disk that forms during the early phase of protostellar collapse, and the power-law description is only transient and appropriate for at most a few $10^4$ kyr. 

\section{Discussion}
In this section, we discuss the limitations of our initial conditions and justify the absence of physical effects such as magnetic fields in our model setup. Moreover, we compare our results with observations of structures around young stellar objects, and we discuss the impact of late encounter events on the formation process of stars, their disks, and eventually planets.

\subsection{Initial conditions}
To consistently carry out a feasible parameter study, we adopted a rather simplified setup.
The initial conditions of our models are idealized in the way that we adopt a two-phase medium consisting of a perfectly spherical cloudlet and a smooth background medium. As described in the introduction GMCs are threaded by filaments, and hence in reality the GMC is not as smooth as considered in our setup. Nevertheless, gas condensations of higher mass similar to our cloudlets likely exist though their shape may deviate significantly from a perfect sphere. Considering the star formation process in a GMC, the analysis of \cite{Jensen2018} shows that some stars undergo late accretion events that cause an increase in luminosity $>1$ Myr after their birth. Our scenario is idealized, but observations of luminosity bursts \citep{Kenyon1990,Evans2009,Dunham2010} indicate that late infall, and hence late encounter events happen for at least some stars. Testing the frequency of such events requires the analysis of star formation simulations, in which the evolution of the stars has evolved with sufficient resolution for a few Myr. 
Along that line, we point out that the main effect of mass replenishment is independent from the shape of the captured or infalling cloudlet as long as the gas encounters the star closely enough. Considering the formation of disks and their associated larger-scale arc-structures, we also expect to see disk formation as a consequence of angular momentum conservation for realistic elongated gas condensations as long as the impact parameter is small enough that mass can be stripped off.

\subsection{Additional physical effects: accretion, irradiation and magnetic fields}
Another simplification of our initial condition is the approximation of a star as a point of constant mass. In reality, a young star accretes mass and it acts as a strong luminosity source and effects its surrounding via irradiation. Although, gas can accrete onto the central mass if it fulfills the accretion criterion of being gravitationally bound and close enough to the sink, our recipe is very simple compared to more expensive simulations investigating the statistical properties of star formation \citep[e.g.][]{Federrath2011, Greif2011, Haugboelle2018}. 
The recipe leads to a drop of the inner density profile, and we tested that for a larger accretion region of 50 au the drop of the surface density profile consistently occurs at a larger radii. To get a more accurate profile at smaller distances would be possible by using a smaller softening radius and smaller accretion radius. However, as the computational time step scales with the Keplerian speed, the simulation time would become too long for our purposes. 
Previous models have shown that late infall events correlate with enhanced accretion rates of up to $\sim10^{-5}$ M$_{\odot}$yr$^{-1}$ over timescales of $\Delta t \sim10^3$ yr \citep{Baraffe2012,Padoan2014}. However, in this study we are interested in the formation of second-generation disks and structures on scales of $\sim1000$ au, rather than in the accretion rate of the already formed host star.
To obtain more realistic accretion rates would require to run a global simulation covering a larger range of scales and incorporating corresponding effects present in the Giant Molecular Cloud \citep[e.g.][]{Haugboelle2018,Bate2018,Wurster2019}. 
Accretion also enhances the luminosity of the host star due to $L\propto \dot{M}$, and can hence cause radiation. 
In some of our runs, we account for stellar irradiation as a constant parameter and find that replenishment of gas in the stellar surrounding as well as the formation of second generation disks is at most barely affected by stellar irradiation. Certainly, our temperature relation $T(r)$ that is purely dependent on radius and constant in time is very simplified, but given that we assume a rather strong luminosity of $50 L_{\odot}$ and no shielding, we rather over- than underestimate irradiation from the central star in our cases. For more realistic setups, one would also need to account for the effects of stellar irradiation from background and neighboring stars, which would affect the thermal pressure in the cloudlet and the forming disk.  

In our models, we do not account for the presence of magnetic fields in GMCs. Adding magnetic fields to our setup would provide an efficient way of transporting angular momentum from the freshly forming disk via magnetic braking \citep{MestelSpitzer1956}. Hence, we expect that magnetic fields would allow the gas to more densely populate the inner parts of the disks ($r\lesssim 100$ au) than seen in our models, and also yield smaller disk sizes. 

Certainly, constraining these additional effects is important and interesting, but beyond the purpose of this paper.
Our aim is to constrain the properties of encounter events rather than carrying out a comprehensive zoom-in simulations of star and disk formation \citep{Kuffmeier2017,Kuffmeier2018,Bate2018,Wurster2019}. By carrying out these more simplified case studies, we have the power to systematically vary the key parameters, and hence to understand the underlying processes.
Nevertheless, to make further progress in understanding capture/infall events, more advanced models are required that provide better statistics about the probabilities and properties of such encounter events. 

\subsection{Comparison with observations}
The presence of a large bent arm around AB Aurigae is a prominent candidate for the outcome of a late encounter event \citep{NakajimaGolimowski1995}. Our models show that disks are typically a few to several 100 au in radius and the large-scale arc-like structures typically extend to distances of several 1000 au from the second-generation disk. The combination of a disk size of at least 430 au \citep{Fukagawa2004}, the large arc-shaped reflection nebula ranging from $\sim$1300 au to $\sim$6000 au \citep{Grady1999} and the age of AB Aurigae, between $2$ Myr \citep{vandenAncker1998} and 4 Myr \citep{DeWarf2003}, is intriguingly consistent with the results of our scenario. Moreover, our study shows the frequent formation of spiral arms inside second-generation disks, which is consistent with scattered light observations of a spiral arm in the disk around AB Aurigae \citep{Grady1999,Fukagawa2004}.

Another candidate for a late infall event of a cloudlet is the 10 Myr \citep{vandenAncker1998} old star HD 1000546 \citep{Grady2001,Ardila2007} with a disk ranging to about 300 au and an arc-shaped envelope extending to about 1000 au from the south-west part of the disk. 
As discussed in depth in D19, arc-shaped structures seem to be more common around Herbig stars than lower mass stars, which might be a consequence of the larger gravitational potential of Herbig stars corresponding to more pronounced arcs or streamers.

Although AB Aurigae and HD 100546 are the most prominent candidate for an ongoing cloudlet encounter event, other Herbig Ae stars may have undergone encounter events in their past, too.
In this context, we want to raise attention to the effect of the background flow on the large-scale arcs.
At the end of the simulations (corresponding to about $10^5$ yrs) with background velocity, the large-scale arc-shaped structures have been or are about to be swept away by the background gas, while a second-generation disks can still be present. 
Therefore, some disks around older stars may in fact be the result of a late encounter event, and thus of second-generation.

Apart from the possibility of encounter events after dispersal of the primordial disk considered in this work, similar encounter events may well happen at earlier stages of the star formation process. For instance, in recent magnetohydrodynamical zoom-in simulations \cite{Kuffmeier2019} find analogous to \cite{Offner2010} that protostellar companions form with initially wider separation and subsequently approach each other, thereby also replenishing the mass reservoir of the primary stars. Prominent candidates for an encounter event happening at earlier times are FU Orionis and Z CMa. Both sources show the presence of bent arms of several 1000 au in size connected to them \citep{NakajimaGolimowski1995,Liu2016,Liu2018}, and both stars experienced episodic accretion events in the past. Therefore, the concept of infall of fresh gas after the initial protostellar collapse phase seems to apply at earlier stages already. 

Considering encounter events at a stage, when the primordial disk is still present raises another interesting question. Given that the second-generation disks in our study typically have lower densities at radii less than 100 au, infall of material with different angular momentum might be an alternative explanation for misalignment of inner and outer disks as observed for HD 142527 \citep{Avenhaus2014,Marino2015}, HD 135344B \citep{Stolker2016} or HD 100453 \citep{Benisty2017}. We investigate the effect of infall events similar to the one studied in this manuscript, but with a primordial disk in an upcoming paper.

Moreover, the density distribution in the second-generation disks correlates with a pressure bump at large radii. As a consequence, we expect dust to pile up in the outer part of the disk \citep{Whipple1972,Brauer2008} opening a possible path for planet formation at large radii in these second-generation disks if enough mass is provided by late-encounter event.

\section{Summary and conclusion}
In this paper, we investigated multiple scenarios of a cloudlet flying by a star that has finished its early phase of formation.  We carried out a parameter study with the moving-mesh code \arepo\ consisting of in total 24 three-dimensional hydrodynamical simulations. In our standard runs, we consider a 2.5 $M_{\odot}$ star that travels through the interstellar medium and encounters a cloudlet with varying size $R_{\rm cloudlet}$ and impact parameter $b$. Building up on these standard runs (capturing scenario), we then consistently analyze the effect of adding turbulence to the initial cloudlet setup, as well as we consider the scenario, where only the cloudlet is moving toward the star, while the background gas and the star are at rest (infall scenario). 
All of these runs are carried out for two extreme cases of perfect cooling (isothermal setup) and no cooling (adiabatic run).
In the final step, we also investigate a capturing scenario in which we account for stellar irradiation by adopting a radially dependent temperature profile rather than a pure isothermal or adiabatic setup.

Due to substantial thermal pressure support, we find no or only modest replenishment of gas during an encounter event for the adiabatic runs consistent with previous results. 
In contrast, for all isothermal runs, we find 
that a substantial amount of mass is captured by the star at distances of up to $\sim 1000$ au from the star. 
These results are in good agreement with previous results by D19 using the \pluto\ code.
Accounting for stellar irradiation does not hinder capturing of the gas and the additional thermal pressure at most mildly reduces the infall onto the central potential. We therefore conclude that stellar irradiation only affects the capturing process on radial distances of $\sim$10 au or less, but not the overall presence of rotational structures such as arcs and disks.
Adding turbulence to the cloudlet only causes very mild deviations.

By following the amount of mass that is enclosed within a radial distance of 1000 au from the star ($M_{\rm enc}$), we find a characteristic sequence of the encounter event for smaller cloudlet radii and substantial impact parameters $b$. 
First, the cloudlet flies by the star leading to a local maximum in the amount of enclosed mass. Afterwards, part of the passing mass returns toward the star from the opposite direction due to the gravitational force of the star. 
Without background velocity (infall scenario), $M_{\rm enc}$ continuously increases due to infall of the background gas. With background velocity (capturing scenario) $M_{\rm enc}$ decreases after a second local maximum because the background gas is swept away by the background flow, and hence there is not enough infall to compensate for the amount of gas accreting onto the star. 

We find that an encounter event quickly leads to the formation of large disks with sizes of several 100 au for runs accounting for cooling, i.e. the isothermal and stellar irradiation runs. The impact parameter $b$ and the cloudlet radius $R_{\rm cloudlet}$, and hence the angular momentum of the cloudlet determine both the inner and outer radial extent of the Keplerian profile of the disks. 
The azimuthal velocity profile tapers off more softly in the infall scenario than in the capturing scenario. In the capturing scenario, the background wind sweeps away the low-density gas leading to a steep drop in the rotational velocity profile beyond the radial extent of the disk ($r\sim100$ au to $1000$ au).

Similar to observations around Herbig stars, we find the presence of arc-structures extending from the outer parts of the disk to distances of about $10^4$ au. These arc structures are bent and longer preserved in the infall scenario. 
Consistent with angular momentum conservation, we find that the extent of rotational structures such as spiral arms, arcs and disks decreases with decreasing impact parameter $b$. For identical impact parameter $b$, but larger cloudlet radius $R_{\rm cloudlet}$, less mass is captured and accretes onto the star relative to the initial cloudlet mass. 

The column density profiles of the disks initially follows a column density distribution that is roughly consistent with a slope of $\Sigma \propto r^{-\frac{3}{2}}$, and afterwards develops toward a profile with a bump at a few $\sim$ 100 au. Toward the end of the sequence the bump in the column density migrates toward smaller radii (about $100$ au) and starts shrinking again. The evolution is correlated with the formation and evolution of a spiral structure that becomes increasingly narrow and dense over time.

Our results show how star-cloudlet encounters can replenish the mass reservoir around an already formed star. 
Furthermore, the results demonstrate that arc structures observed for AB Aurigae or HD 100546 are a likely consequence of such late encounter events. We find that large second-generation disks can form via encounter events of a star with denser gas condensations in the ISM millions of years after stellar birth as long as the parental Giant Molecular Cloud has not fully dispersed. The majority of mass in these second-generation disks is located at large radii, which is consistent with observations of transitional disks. 

\begin{acknowledgements}
The research of MK is supported by a research grant of the Independent Research Foundation Denmark (IRFD) (international postdoctoral fellow, project number: 8028-00025B). 
MK acknowledges the support of the DFG Research Unit `Transition Disks' (FOR 2634/1, DU 414/23-1). 
FGG acknowledges support from DFG Schwerpunktprogramm SPP 1992 ``Diversity of Extrasolar Planets'' grant DU414/21-1.
Last but not least, we thank the referee for his constructive comments and suggestions on an earlier draft of the manuscript.
\end{acknowledgements}

%-------------------------------------------------------------------

\bibliography{general}

\begin{thebibliography}{73}
\expandafter\ifx\csname natexlab\endcsname\relax\def\natexlab#1{#1}\fi

\bibitem[{{Ardila} {et~al.}(2007){Ardila}, {Golimowski}, {Krist}, {Clampin},
  {Ford}, \& {Illingworth}}]{Ardila2007}
{Ardila}, D.~R., {Golimowski}, D.~A., {Krist}, J.~E., {et~al.} 2007, \apj, 665,
  512

\bibitem[{{Avenhaus} {et~al.}(2014){Avenhaus}, {Quanz}, {Schmid}, {Meyer},
  {Garufi}, {Wolf}, \& {Dominik}}]{Avenhaus2014}
{Avenhaus}, H., {Quanz}, S.~P., {Schmid}, H.~M., {et~al.} 2014, \apj, 781, 87

\bibitem[{{Baraffe} {et~al.}(2009){Baraffe}, {Chabrier}, \&
  {Gallardo}}]{Baraffe2009}
{Baraffe}, I., {Chabrier}, G., \& {Gallardo}, J. 2009, \apjl, 702, L27

\bibitem[{{Baraffe} {et~al.}(2012){Baraffe}, {Vorobyov}, \&
  {Chabrier}}]{Baraffe2012}
{Baraffe}, I., {Vorobyov}, E., \& {Chabrier}, G. 2012, \apj, 756, 118

\bibitem[{{Bate}(2018)}]{Bate2018}
{Bate}, M.~R. 2018, \mnras, 475, 5618

\bibitem[{{Benisty} {et~al.}(2017){Benisty}, {Stolker}, {Pohl}, {de Boer},
  {Lesur}, {Dominik}, {Dullemond}, {Langlois}, {Min}, {Wagner}, {Henning},
  {Juhasz}, {Pinilla}, {Facchini}, {Apai}, {van Boekel}, {Garufi}, {Ginski},
  {M{\'e}nard}, {Pinte}, {Quanz}, {Zurlo}, {Boccaletti}, {Bonnefoy}, {Beuzit},
  {Chauvin}, {Cudel}, {Desidera}, {Feldt}, {Fontanive}, {Gratton}, {Kasper},
  {Lagrange}, {LeCoroller}, {Mouillet}, {Mesa}, {Sissa}, {Vigan}, {Antichi},
  {Buey}, {Fusco}, {Gisler}, {Llored}, {Magnard}, {Moeller-Nilsson}, {Pragt},
  {Roelfsema}, {Sauvage}, \& {Wildi}}]{Benisty2017}
{Benisty}, M., {Stolker}, T., {Pohl}, A., {et~al.} 2017, \aap, 597, A42

\bibitem[{{Blitz}(1993)}]{Blitz1993}
{Blitz}, L. 1993, in Protostars and Planets III, ed. E.~H. {Levy} \& J.~I.
  {Lunine}, 125--161

\bibitem[{{Bondi}(1952)}]{Bondi1952}
{Bondi}, H. 1952, \mnras, 112, 195

\bibitem[{{Bondi} \& {Hoyle}(1944)}]{BondiHoyle1944}
{Bondi}, H. \& {Hoyle}, F. 1944, \mnras, 104, 273

\bibitem[{{Brauer} {et~al.}(2008){Brauer}, {Henning}, \&
  {Dullemond}}]{Brauer2008}
{Brauer}, F., {Henning}, T., \& {Dullemond}, C.~P. 2008, \aap, 487, L1

\bibitem[{{DeWarf} {et~al.}(2003){DeWarf}, {Sepinsky}, {Guinan}, {Ribas}, \&
  {Nadalin}}]{DeWarf2003}
{DeWarf}, L.~E., {Sepinsky}, J.~F., {Guinan}, E.~F., {Ribas}, I., \& {Nadalin},
  I. 2003, \apj, 590, 357

\bibitem[{{Dubinski} {et~al.}(1995){Dubinski}, {Narayan}, \&
  {Phillips}}]{Dubinski1995}
{Dubinski}, J., {Narayan}, R., \& {Phillips}, T.~G. 1995, \apj, 448, 226

\bibitem[{{Dullemond} {et~al.}(2019){Dullemond}, {K{\"u}ffmeier}, {Goicovic},
  {Fukagawa}, {Oehl}, \& {Kramer}}]{Dullemond2019}
{Dullemond}, C.~P., {K{\"u}ffmeier}, M., {Goicovic}, F., {et~al.} 2019, \aap,
  628, A20

\bibitem[{{Dunham} {et~al.}(2010){Dunham}, {Evans}, {Terebey}, {Dullemond}, \&
  {Young}}]{Dunham2010}
{Dunham}, M.~M., {Evans}, II, N.~J., {Terebey}, S., {Dullemond}, C.~P., \&
  {Young}, C.~H. 2010, \apj, 710, 470

\bibitem[{{Evans} {et~al.}(2009){Evans}, {Dunham}, {J{\o}rgensen}, {Enoch},
  {Mer{\'{\i}}n}, {van Dishoeck}, {Alcal{\'a}}, {Myers}, {Stapelfeldt},
  {Huard}, {Allen}, {Harvey}, {van Kempen}, {Blake}, {Koerner}, {Mundy},
  {Padgett}, \& {Sargent}}]{Evans2009}
{Evans}, II, N.~J., {Dunham}, M.~M., {J{\o}rgensen}, J.~K., {et~al.} 2009,
  \apjs, 181, 321

\bibitem[{{Falgarone} {et~al.}(2004){Falgarone}, {Hily-Blant}, \&
  {Levrier}}]{Falgarone2004}
{Falgarone}, E., {Hily-Blant}, P., \& {Levrier}, F. 2004, \apss, 292, 89

\bibitem[{{Federrath} {et~al.}(2011){Federrath}, {Banerjee}, {Seifried},
  {Clark}, \& {Klessen}}]{Federrath2011}
{Federrath}, C., {Banerjee}, R., {Seifried}, D., {Clark}, P.~C., \& {Klessen},
  R.~S. 2011, in IAU Symposium, Vol. 270, Computational Star Formation, ed.
  J.~{Alves}, B.~G. {Elmegreen}, J.~M. {Girart}, \& V.~{Trimble}, 425--428

\bibitem[{{Field} {et~al.}(1969){Field}, {Goldsmith}, \& {Habing}}]{Field1969}
{Field}, G.~B., {Goldsmith}, D.~W., \& {Habing}, H.~J. 1969, \apjl, 155, L149

\bibitem[{{Frimann} {et~al.}(2017){Frimann}, {J{\o}rgensen}, {Dunham},
  {Bourke}, {Kristensen}, {Offner}, {Stephens}, {Tobin}, \&
  {Vorobyov}}]{Frimann2017}
{Frimann}, S., {J{\o}rgensen}, J.~K., {Dunham}, M.~M., {et~al.} 2017, \aap,
  602, A120

\bibitem[{{Fukagawa} {et~al.}(2004){Fukagawa}, {Hayashi}, {Tamura}, {Itoh},
  {Hayashi}, {Oasa}, {Takeuchi}, {Morino}, {Murakawa}, {Oya}, {Yamashita},
  {Suto}, {Mayama}, {Naoi}, {Ishii}, {Pyo}, {Nishikawa}, {Takato}, {Usuda},
  {Ando}, {Iye}, {Miyama}, \& {Kaifu}}]{Fukagawa2004}
{Fukagawa}, M., {Hayashi}, M., {Tamura}, M., {et~al.} 2004, \apj, 605, L53

\bibitem[{{Goicovic} {et~al.}(2019){Goicovic}, {Springel}, {Ohlmann}, \&
  {Pakmor}}]{Goicovic2019}
{Goicovic}, F.~G., {Springel}, V., {Ohlmann}, S.~T., \& {Pakmor}, R. 2019,
  \mnras, 487, 981

\bibitem[{{Grady} {et~al.}(2001){Grady}, {Polomski}, {Henning}, {Stecklum},
  {Woodgate}, {Telesco}, {Pi{\~n}a}, {Gull}, {Boggess}, {Bowers}, {Bruhweiler},
  {Clampin}, {Danks}, {Green}, {Heap}, {Hutchings}, {Jenkins}, {Joseph},
  {Kaiser}, {Kimble}, {Kraemer}, {Lindler}, {Linsky}, {Maran}, {Moos}, {Plait},
  {Roesler}, {Timothy}, \& {Weistrop}}]{Grady2001}
{Grady}, C.~A., {Polomski}, E.~F., {Henning}, T., {et~al.} 2001, \aj, 122, 3396

\bibitem[{{Grady} {et~al.}(1999){Grady}, {Woodgate}, {Bruhweiler}, {Boggess},
  {Plait}, {Lindler}, {Clampin}, \& {Kalas}}]{Grady1999}
{Grady}, C.~A., {Woodgate}, B., {Bruhweiler}, F.~C., {et~al.} 1999, \apj, 523,
  L151

\bibitem[{{Greif} {et~al.}(2011){Greif}, {Springel}, {White}, {Glover},
  {Clark}, {Smith}, {Klessen}, \& {Bromm}}]{Greif2011}
{Greif}, T.~H., {Springel}, V., {White}, S. D.~M., {et~al.} 2011, \apj, 737, 75

\bibitem[{{Haisch} {et~al.}(2001){Haisch}, {Lada}, \& {Lada}}]{Haisch2001}
{Haisch}, Jr., K.~E., {Lada}, E.~A., \& {Lada}, C.~J. 2001, \apjl, 553, L153

\bibitem[{{Haugb{\o}lle} {et~al.}(2018){Haugb{\o}lle}, {Padoan}, \&
  {Nordlund}}]{Haugboelle2018}
{Haugb{\o}lle}, T., {Padoan}, P., \& {Nordlund}, {\AA}. 2018, \apj, 854, 35

\bibitem[{{Heiles}(1997)}]{Heiles1997}
{Heiles}, C. 1997, \apj, 481, 193

\bibitem[{{Hoyle} \& {Lyttleton}(1939)}]{HoyleLyttleton1939}
{Hoyle}, F. \& {Lyttleton}, R.~A. 1939, Proceedings of the Cambridge
  Philosophical Society, 35, 405

\bibitem[{{Jensen} \& {Haugb{\o}lle}(2018)}]{Jensen2018}
{Jensen}, S.~S. \& {Haugb{\o}lle}, T. 2018, \mnras, 474, 1176

\bibitem[{{J{\o}rgensen} {et~al.}(2015){J{\o}rgensen}, {Visser}, {Williams}, \&
  {Bergin}}]{Jorgensen2015}
{J{\o}rgensen}, J.~K., {Visser}, R., {Williams}, J.~P., \& {Bergin}, E.~A.
  2015, \aap, 579, A23

\bibitem[{{Kenyon} {et~al.}(1990){Kenyon}, {Hartmann}, {Strom}, \&
  {Strom}}]{Kenyon1990}
{Kenyon}, S.~J., {Hartmann}, L.~W., {Strom}, K.~M., \& {Strom}, S.~E. 1990,
  \aj, 99, 869

\bibitem[{{Keppler} {et~al.}(2018){Keppler}, {Benisty}, {M{\"u}ller},
  {Henning}, {van Boekel}, {Cantalloube}, {Ginski}, {van Holstein}, {Maire},
  {Pohl}, {Samland}, {Avenhaus}, {Baudino}, {Boccaletti}, {de Boer},
  {Bonnefoy}, {Chauvin}, {Desidera}, {Langlois}, {Lazzoni}, {Marleau},
  {Mordasini}, {Pawellek}, {Stolker}, {Vigan}, {Zurlo}, {Birnstiel},
  {Brandner}, {Feldt}, {Flock}, {Girard}, {Gratton}, {Hagelberg}, {Isella},
  {Janson}, {Juhasz}, {Kemmer}, {Kral}, {Lagrange}, {Launhardt}, {Matter},
  {M{\'e}nard}, {Milli}, {Molli{\`e}re}, {Olofsson}, {P{\'e}rez}, {Pinilla},
  {Pinte}, {Quanz}, {Schmidt}, {Udry}, {Wahhaj}, {Williams}, {Buenzli},
  {Cudel}, {Dominik}, {Galicher}, {Kasper}, {Lannier}, {Mesa}, {Mouillet},
  {Peretti}, {Perrot}, {Salter}, {Sissa}, {Wildi}, {Abe}, {Antichi},
  {Augereau}, {Baruffolo}, {Baudoz}, {Bazzon}, {Beuzit}, {Blanchard}, {Brems},
  {Buey}, {De Caprio}, {Carbillet}, {Carle}, {Cascone}, {Cheetham}, {Claudi},
  {Costille}, {Delboulb{\'e}}, {Dohlen}, {Fantinel}, {Feautrier}, {Fusco},
  {Giro}, {Gluck}, {Gry}, {Hubin}, {Hugot}, {Jaquet}, {Le Mignant}, {Llored},
  {Madec}, {Magnard}, {Martinez}, {Maurel}, {Meyer}, {M{\"o}ller-Nilsson},
  {Moulin}, {Mugnier}, {Orign{\'e}}, {Pavlov}, {Perret}, {Petit}, {Pragt},
  {Puget}, {Rabou}, {Ramos}, {Rigal}, {Rochat}, {Roelfsema}, {Rousset}, {Roux},
  {Salasnich}, {Sauvage}, {Sevin}, {Soenke}, {Stadler}, {Suarez}, {Turatto}, \&
  {Weber}}]{Keppler2018}
{Keppler}, M., {Benisty}, M., {M{\"u}ller}, A., {et~al.} 2018, \aap, 617, A44

\bibitem[{{Klessen} \& {Hennebelle}(2010)}]{KlessenHennebelle2010}
{Klessen}, R.~S. \& {Hennebelle}, P. 2010, \aap, 520, A17

\bibitem[{{Kuffmeier} {et~al.}(2019){Kuffmeier}, {Calcutt}, \&
  {Kristensen}}]{Kuffmeier2019}
{Kuffmeier}, M., {Calcutt}, H., \& {Kristensen}, L.~E. 2019, arXiv e-prints,
  arXiv:1907.02083

\bibitem[{{Kuffmeier} {et~al.}(2018){Kuffmeier}, {Frimann}, {Jensen}, \&
  {Haugb{\o}lle}}]{Kuffmeier2018}
{Kuffmeier}, M., {Frimann}, S., {Jensen}, S.~S., \& {Haugb{\o}lle}, T. 2018,
  \mnras, 475, 2642

\bibitem[{{Kuffmeier} {et~al.}(2017){Kuffmeier}, {Haugb{\o}lle}, \&
  {Nordlund}}]{Kuffmeier2017}
{Kuffmeier}, M., {Haugb{\o}lle}, T., \& {Nordlund}, {\AA}. 2017, \apj, 846, 7

\bibitem[{{Kunitomo} {et~al.}(2017){Kunitomo}, {Guillot}, {Takeuchi}, \&
  {Ida}}]{Kunitomo2017}
{Kunitomo}, M., {Guillot}, T., {Takeuchi}, T., \& {Ida}, S. 2017, \aap, 599,
  A49

\bibitem[{{Langer} {et~al.}(1995){Langer}, {Velusamy}, {Kuiper}, {Levin},
  {Olsen}, \& {Migenes}}]{Langer1995}
{Langer}, W.~D., {Velusamy}, T., {Kuiper}, T.~B.~H., {et~al.} 1995, \apj, 453,
  293

\bibitem[{{Larson}(1969)}]{Larson1969}
{Larson}, R.~B. 1969, \mnras, 145, 271

\bibitem[{{Larson}(1981)}]{Larson1981}
{Larson}, R.~B. 1981, \mnras, 194, 809

\bibitem[{{Li} {et~al.}(2011){Li}, {Krasnopolsky}, \& {Shang}}]{Li2011}
{Li}, Z.-Y., {Krasnopolsky}, R., \& {Shang}, H. 2011, \apj, 738, 180

\bibitem[{{Liu} {et~al.}(2018){Liu}, {Dunham}, {Pascucci}, {Bourke}, {Hirano},
  {Longmore}, {Andrews}, {Carrasco-Gonz{\'a}lez}, {Forbrich},
  {Galv{\'a}n-Madrid}, {Girart}, {Green}, {Ju{\'a}rez}, {K{\'o}sp{\'a}l},
  {Manara}, {Palau}, {Takami}, {Testi}, \& {Vorobyov}}]{Liu2018}
{Liu}, H.~B., {Dunham}, M.~M., {Pascucci}, I., {et~al.} 2018, \aap, 612, A54

\bibitem[{{Liu} {et~al.}(2016){Liu}, {Takami}, {Kudo}, {Hashimoto}, {Dong},
  {Vorobyov}, {Pyo}, {Fukagawa}, {Tamura}, {Henning}, {Dunham}, {Karr},
  {Kusakabe}, \& {Tsuribe}}]{Liu2016}
{Liu}, H.~B., {Takami}, M., {Kudo}, T., {et~al.} 2016, Science Advances, 2,
  e1500875

\bibitem[{{Mac Low} \& {Klessen}(2004)}]{MacLowKlessen2004}
{Mac Low}, M.-M. \& {Klessen}, R.~S. 2004, Reviews of Modern Physics, 76, 125

\bibitem[{{Machida} {et~al.}(2010){Machida}, {Inutsuka}, \&
  {Matsumoto}}]{Machida2010}
{Machida}, M.~N., {Inutsuka}, S.-i., \& {Matsumoto}, T. 2010, \apj, 724, 1006

\bibitem[{{Machida} \& {Matsumoto}(2011)}]{Machida2011}
{Machida}, M.~N. \& {Matsumoto}, T. 2011, \mnras, 413, 2767

\bibitem[{{Mamajek}(2009)}]{Mamajek2009}
{Mamajek}, E.~E. 2009, in American Institute of Physics Conference Series, Vol.
  1158, American Institute of Physics Conference Series, ed. T.~{Usuda},
  M.~{Tamura}, \& M.~{Ishii}, 3--10

\bibitem[{{Manara} {et~al.}(2018){Manara}, {Morbidelli}, \&
  {Guillot}}]{Manara2018}
{Manara}, C.~F., {Morbidelli}, A., \& {Guillot}, T. 2018, \aap, 618, L3

\bibitem[{{Marino} {et~al.}(2015){Marino}, {Perez}, \& {Casassus}}]{Marino2015}
{Marino}, S., {Perez}, S., \& {Casassus}, S. 2015, \apjl, 798, L44

\bibitem[{{Masson} {et~al.}(2016){Masson}, {Chabrier}, {Hennebelle}, {Vaytet},
  \& {Commer{\c c}on}}]{Masson2016}
{Masson}, J., {Chabrier}, G., {Hennebelle}, P., {Vaytet}, N., \& {Commer{\c
  c}on}, B. 2016, \aap, 587, A32

\bibitem[{{Mestel} \& {Spitzer}(1956)}]{MestelSpitzer1956}
{Mestel}, L. \& {Spitzer}, Jr., L. 1956, \mnras, 116, 503

\bibitem[{{Mignone} {et~al.}(2007){Mignone}, {Bodo}, {Massaglia}, {Matsakos},
  {Tesileanu}, {Zanni}, \& {Ferrari}}]{Mignone2007}
{Mignone}, A., {Bodo}, G., {Massaglia}, S., {et~al.} 2007, \apjs, 170, 228

\bibitem[{{Moeckel} \& {Throop}(2009)}]{MoeckelThroop2009}
{Moeckel}, N. \& {Throop}, H.~B. 2009, \apj, 707, 268

\bibitem[{{Mulders} {et~al.}(2015){Mulders}, {Pascucci}, \&
  {Apai}}]{Mulders2015}
{Mulders}, G.~D., {Pascucci}, I., \& {Apai}, D. 2015, \apj, 814, 130

\bibitem[{{Nakajima} \& {Golimowski}(1995)}]{NakajimaGolimowski1995}
{Nakajima}, T. \& {Golimowski}, D.~A. 1995, \aj, 109, 1181

\bibitem[{{Nixon} {et~al.}(2018){Nixon}, {King}, \& {Pringle}}]{Nixon2018}
{Nixon}, C.~J., {King}, A.~R., \& {Pringle}, J.~E. 2018, \mnras, 477, 3273

\bibitem[{{Offner} {et~al.}(2010){Offner}, {Kratter}, {Matzner}, {Krumholz}, \&
  {Klein}}]{Offner2010}
{Offner}, S.~S.~R., {Kratter}, K.~M., {Matzner}, C.~D., {Krumholz}, M.~R., \&
  {Klein}, R.~I. 2010, \apj, 725, 1485

\bibitem[{{Padoan} {et~al.}(2014){Padoan}, {Haugb{\o}lle}, \&
  {Nordlund}}]{Padoan2014}
{Padoan}, P., {Haugb{\o}lle}, T., \& {Nordlund}, {\AA}. 2014, \apj, 797, 32

\bibitem[{{Padoan} \& {Nordlund}(2002)}]{Padoan_turbfrag}
{Padoan}, P. \& {Nordlund}, {\AA}. 2002, \apj, 576, 870

\bibitem[{{Pakmor} {et~al.}(2016){Pakmor}, {Springel}, {Bauer}, {Mocz},
  {Munoz}, {Ohlmann}, {Schaal}, \& {Zhu}}]{Pakmor2016}
{Pakmor}, R., {Springel}, V., {Bauer}, A., {et~al.} 2016, \mnras, 455, 1134

\bibitem[{{Seifried} {et~al.}(2013){Seifried}, {Banerjee}, {Pudritz}, \&
  {Klessen}}]{Seifried2013}
{Seifried}, D., {Banerjee}, R., {Pudritz}, R.~E., \& {Klessen}, R.~S. 2013,
  \mnras, 432, 3320

\bibitem[{{Springel}(2010)}]{Springel2010}
{Springel}, V. 2010, \mnras, 401, 791

\bibitem[{{Stolker} {et~al.}(2016){Stolker}, {Dominik}, {Avenhaus}, {Min}, {de
  Boer}, {Ginski}, {Schmid}, {Juhasz}, {Bazzon}, {Waters}, {Garufi},
  {Augereau}, {Benisty}, {Boccaletti}, {Henning}, {Langlois}, {Maire},
  {M{\'e}nard}, {Meyer}, {Pinte}, {Quanz}, {Thalmann}, {Beuzit}, {Carbillet},
  {Costille}, {Dohlen}, {Feldt}, {Gisler}, {Mouillet}, {Pavlov}, {Perret},
  {Petit}, {Pragt}, {Rochat}, {Roelfsema}, {Salasnich}, {Soenke}, \&
  {Wildi}}]{Stolker2016}
{Stolker}, T., {Dominik}, C., {Avenhaus}, H., {et~al.} 2016, \aap, 595, A113

\bibitem[{{Tannirkulam} {et~al.}(2008){Tannirkulam}, {Monnier}, {Harries},
  {Millan-Gabet}, {Zhu}, {Pedretti}, {Ireland}, {Tuthill}, {ten Brummelaar},
  {McAlister}, {Farrington}, {Goldfinger}, {Sturmann}, {Sturmann}, \&
  {Turner}}]{Tannirkulam2008}
{Tannirkulam}, A., {Monnier}, J.~D., {Harries}, T.~J., {et~al.} 2008, \apj,
  689, 513

\bibitem[{{Tomida} {et~al.}(2015){Tomida}, {Okuzumi}, \&
  {Machida}}]{Tomida2015}
{Tomida}, K., {Okuzumi}, S., \& {Machida}, M.~N. 2015, \apj, 801, 117

\bibitem[{{van den Ancker} {et~al.}(1998){van den Ancker}, {de Winter}, \&
  {Tjin A Djie}}]{vandenAncker1998}
{van den Ancker}, M.~E., {de Winter}, D., \& {Tjin A Djie}, H.~R.~E. 1998,
  \aap, 330, 145

\bibitem[{{Vorobyov} \& {Basu}(2010)}]{Vorobyov2010}
{Vorobyov}, E.~I. \& {Basu}, S. 2010, \apj, 719, 1896

\bibitem[{{Whipple}(1972)}]{Whipple1972}
{Whipple}, F.~L. 1972, in From Plasma to Planet, ed. A.~{Elvius}, 211

\bibitem[{{Wijnen} {et~al.}(2017{\natexlab{a}}){Wijnen}, {Pelupessy}, {Pols},
  \& {Portegies Zwart}}]{Wijnen2017b}
{Wijnen}, T.~P.~G., {Pelupessy}, F.~I., {Pols}, O.~R., \& {Portegies Zwart}, S.
  2017{\natexlab{a}}, \aap, 604, A88

\bibitem[{{Wijnen} {et~al.}(2016){Wijnen}, {Pols}, {Pelupessy}, \& {Portegies
  Zwart}}]{Wijnen2016}
{Wijnen}, T.~P.~G., {Pols}, O.~R., {Pelupessy}, F.~I., \& {Portegies Zwart}, S.
  2016, \aap, 594, A30

\bibitem[{{Wijnen} {et~al.}(2017{\natexlab{b}}){Wijnen}, {Pols}, {Pelupessy},
  \& {Portegies Zwart}}]{Wijnen2017a}
{Wijnen}, T.~P.~G., {Pols}, O.~R., {Pelupessy}, F.~I., \& {Portegies Zwart}, S.
  2017{\natexlab{b}}, \aap, 602, A52

\bibitem[{{Wurster} {et~al.}(2018){Wurster}, {Bate}, \& {Price}}]{Wurster2018}
{Wurster}, J., {Bate}, M.~R., \& {Price}, D.~J. 2018, \mnras, 475, 1859

\bibitem[{{Wurster} {et~al.}(2019){Wurster}, {Bate}, \& {Price}}]{Wurster2019}
{Wurster}, J., {Bate}, M.~R., \& {Price}, D.~J. 2019, \mnras, 2117

\end{thebibliography}
\bibliographystyle{aa}

%%%%%%%%%%%%%%%%%%%%%%%%%%%%%%%%%%%%%%

\begin{appendix}
\section{Initial grid setup}
\label{app:ic_grid}

The cells in our simulations consist in the high resolution gas of the cloudlet, together with a low resolution medium. Because the latter fills the whole domain, the size difference between the cloudlet's cells and the background can be very large. This creates sharp gradients in the contact areas of this two phases, even when we impose pressure equilibrium.
To avoid spurious shocks induced by these differences, we initialize our background grid such that the volume difference does not go above a certain threshold. For these models we chose a volume difference of 5.

\begin{figure}
    \centering
    \includegraphics[width=\columnwidth]{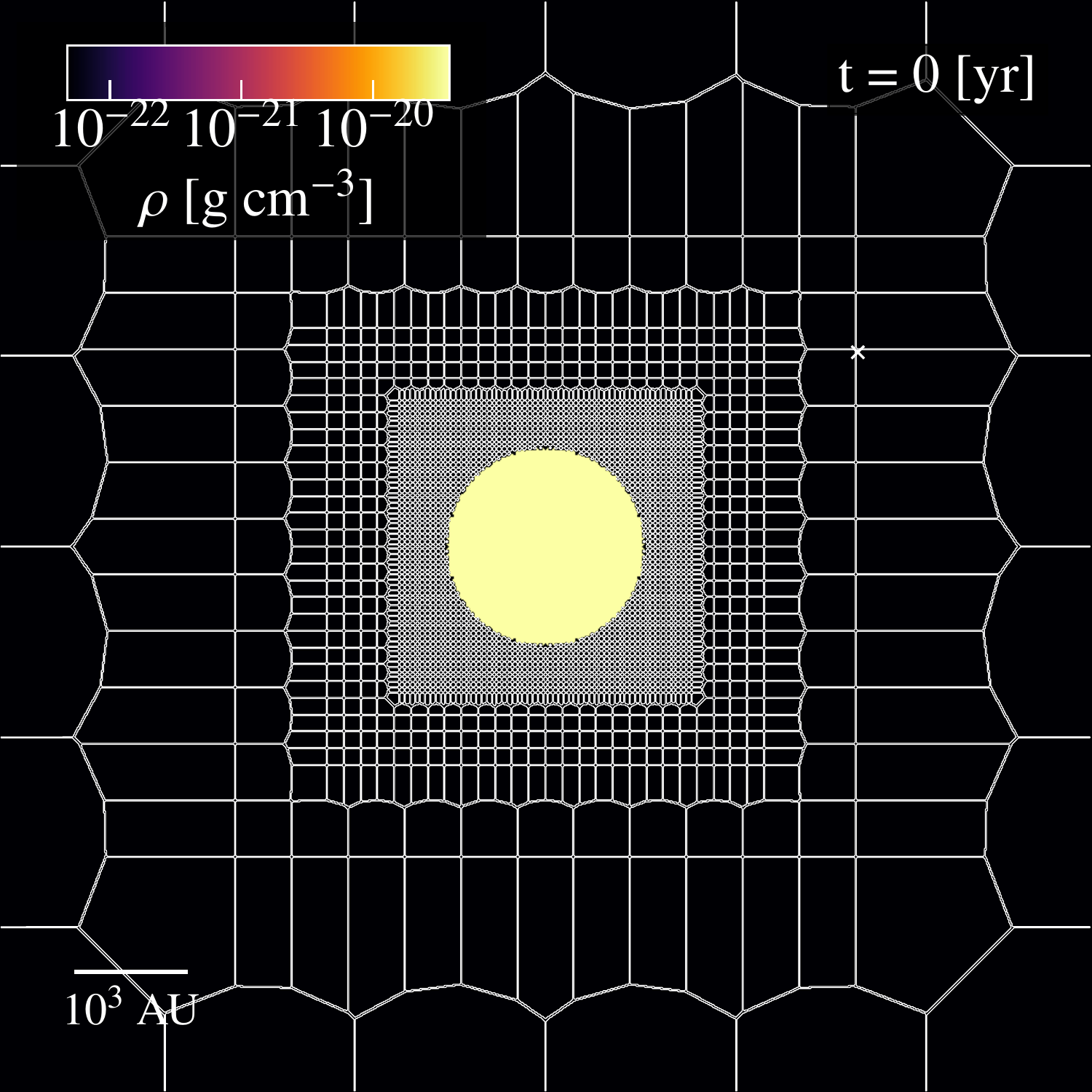}
    \caption{Slice of the grid's initial setup. This illustrate our procedure to guarantee a modest transition between the highly resolved cloud (resolution not shown) and the coarsely resolved background.}
    \label{fig:grid_setup}
\end{figure}

In practice, what we do is to put the cloudlet configuration in a small cubic grid such that the background cells satisfy the volume difference. Then this configuration is surrounded by a slightly larger box under the same condition, and so on until the whole domain is filled with gas.
An illustration of the initial setup is shown in Fig~\ref{fig:grid_setup}.

\section{Refinement around the central potential}
\label{app:refinement}

In order to push the resolution down to scales of a few au close to the star we implemented an additional refinement criterion on top of the standard mass-volume based criteria used for the rest of the gas in the domain.
We define a refinement radius ($r_{\rm refine}$) within which a cell is split if the following conditions are satisfied
\begin{eqnarray}
r_{\rm cell} > \frac{r}{N_{\rm cps}}, &\mbox{if}& r<r_{\rm refine},\\
r_{\rm cell} > \frac{r_{\rm soft}}{N_{\rm cps}}, &\mbox{if}& r<r_{\rm soft},
\end{eqnarray}
where $r_{\rm cell}$ is the cell radius, $r$ is the distance to centre and $N_{\rm csp}$ is the desired number of cells per softening.
For the models presented in this paper we used $r_{\rm refine}=25$ au, and $N_{\rm cps}=5$.

\begin{figure}
    \centering
    \includegraphics[width=\columnwidth]{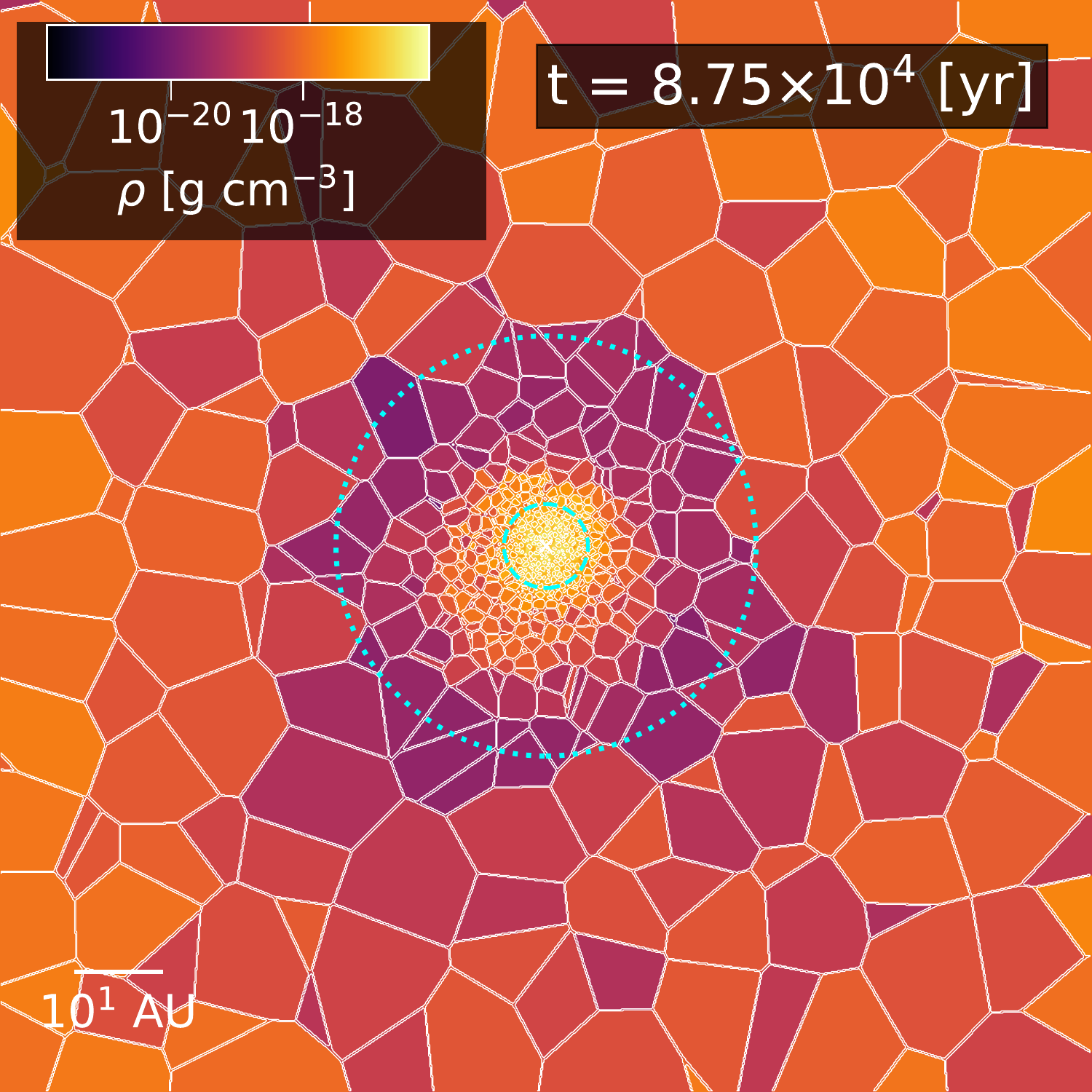}
    \caption{Slice of the gaseous grid close to the central potential. The outer dotted line shows the refinement region defined by $r_{\rm refine}$, while the inner dashed line depicts the softening length of the gravitational potential.}
    \label{fig:grid_refine}
\end{figure}

An example of one of our models with this refinement can be seen in Fig.~\ref{fig:grid_refine}, which depicts a slice through the Voronoi cells around the star. The outer dotted line shows the refinement region $r_{\rm refine}$, while the inner dashed line corresponds to the softening of the gravitational potential $r_{\rm soft}$.
With this procedure we have enough resolution elements with respect to the softening to ensure the numerical stability of the system, and to better resolve the gas dynamics close to the star.
Since we apply such high resolution only in a small region of the whole domain, the computational costs of our simulations only modestly increase.

\end{appendix}

\end{document}